\documentclass[twocolumn]{aastex631}
\DeclareUnicodeCharacter{02BC}{'}
\usepackage{threeparttable}

\usepackage{tabularx}
\usepackage{array}
\usepackage{xcolor}

\newcolumntype{P}[1]{>{\centering\raggedright\arraybackslash}p{#1}}
\newcolumntype{Q}[1]{>{\centering\arraybackslash}p{#1}}
\definecolor{AASBlue}{rgb}{0.0, 0.0, 0.5}

\accepted{July 25, 2025}
\submitjournal{ApJ}

\begin{document}

\title{Jet Collimation Profile of Low-Luminosity AGN M84: Insight into the Jet Formation in the Low Accretion Regime}

\correspondingauthor{Elika Prameswari Fariyanto, \\ Kazuhiro Hada}
\email{fariyanto-elika-prameswari@g.ecc.u-tokyo.ac.jp, hada@nsc.nagoya-cu.ac.jp}

\author[0009-0005-4034-1373]{Elika P. Fariyanto}
\affiliation{Department of Astronomy, Graduate School of Science, The University of Tokyo, 7-3-1 Hongo, Bunkyo-ku, Tokyo 113-0033, Japan}
\affiliation{National Astronomical Observatory of Japan, 2-21-1 Osawa, Mitaka, Tokyo 181-8588, Japan}

\author[0000-0001-6906-772X]{Kazuhiro Hada}
\affiliation{Graduate School of Science, Nagoya City University, Yamanohata 1, Mizuho-cho, Mizuho-ku, Nagoya, 467-8501, Aichi, Japan}
\affiliation{Mizusawa VLBI Observatory, National Astronomical Observatory of Japan, 2-12 Hoshigaoka, Mizusawa, Oshu, Iwate 023-0861, Japan}
\affiliation{Department of Astronomical Science, The Graduate University for Advanced Studies (SOKENDAI), 2-21-1 Osawa, Mitaka, Tokyo 181-8588, Japan}

\author[0000-0001-6311-4345]{Yuzhu Cui}
\affiliation{Institute of Astrophysics, Central China Normal University, Wuhan 430079, China}
\affiliation{Research Center for Intelligent Computing Platforms, Zhejiang Laboratory, Hangzhou 311100, Peopleʼs Republic of China}

\author[0000-0003-4058-9000]{Mareki Honma}
\affiliation{Department of Astronomy, Graduate School of Science, The University of Tokyo, 7-3-1 Hongo, Bunkyo-ku, Tokyo 113-0033, Japan}
\affiliation{Mizusawa VLBI Observatory, National Astronomical Observatory of Japan, 2-12 Hoshigaoka, Mizusawa, Oshu, Iwate 023-0861, Japan}

\author[0000-0001-6081-2420]{Masanori Nakamura}
\affiliation{Department of General Science and Education, National Institute of Technology, Hachinohe College, 16-1 Uwanotai, Tamonoki, Hachinohe, Aomori 039-1192, Japan}
\affiliation{Institute of Astronomy and Astrophysics, Academia Sinica, P.O. Box 23-141, Taipei 10617, Taiwan, R.O.C.}

\author[0000-0001-6988-8763]{Keiichi Asada}
\affiliation{Institute of Astronomy and Astrophysics, Academia Sinica, P.O. Box 23-141, Taipei 10617, Taiwan, R.O.C.}

\author{Xuezheng Wang}
\affiliation{Zhejiang Shuren University, Shuren Street, Hangzhou, Zhejiang 310009, China}

\author[0000-0001-7369-3539]{Wu Jiang}
\affiliation{Shanghai Astronomical Observatory, Chinese Academy of Sciences, Shanghai 200030, China}


\begin{abstract}
Recent advancements in high-resolution Very Long Baseline Interferometry (VLBI) have significantly improved our understanding of jet collimation near supermassive black holes in active galactic nuclei (AGNs), particularly in high-power systems. However, the collimation properties of jets in low-luminosity AGNs (LLAGNs) remain poorly explored. In this study, we investigate the jet structure of M\,84, a nearby radio galaxy and a representative LLAGN, to probe jet collimation properties in a low-accretion regime. Utilizing astrometric phase-referencing observations from the Very Long Baseline Array (VLBA), supplemented by archival Very Large Array (VLA) data, we trace the jet geometry of M\,84 over a broad range of scales, from $\sim 10^2$ to $\sim 10^7$ Schwarzschild radii ($r_{\rm s}$). Our analysis reveals a well-defined transition from a semi-parabolic profile, $W(r) \propto r^{0.71}$, to a conical shape, $W(r) \propto r^{1.16}$, occurring at approximately $1.67 \times 10^4\,r_{\rm s}$. This indicates that the M\,84 jet is notably less collimated than those in other known LLAGN sources. Our findings provide new insights into the relationship between jet collimation and accretion rate, offering crucial constraints for jet formation models in LLAGNs.
\end{abstract}

\keywords{\href{http://astrothesaurus.org/uat/16}{Active galactic nuclei (16)} --- \href{http://astrothesaurus.org/uat/1343}{Radio galaxies (1343)} --- \href{http://astrothesaurus.org/uat/1390}{Relativistic jets (1390)} --- \href{http://astrothesaurus.org/uat/1769}{Very long baseline interferometry (1769)} --- \href{http://astrothesaurus.org/uat/994}{Magnetic fields (994)} --- \href{http://astrothesaurus.org/uat/14}{Accretion (14)}}


\section{Introduction} \label{sec:intro}
Astrophysical jets, characterized by their exceptional collimation and immense energy, are among the most striking phenomena in the universe. In active galactic nuclei (AGNs), the formation of relativistic jets is widely believed to be linked to the presence of a central supermassive black hole (SMBH) and its surrounding accretion flow. The underlying mechanism likely involves energy extraction from the spinning black hole via the Blandford-Znajek process~\citep{blandford1977electromagnetic} and/or the inner regions of the accretion disk through magnetohydrodynamic processes~\citep{blandford1982hydromagnetic}. However, detailed physical processes of jet formation, collimation, and acceleration are still not well understood. 
State-of-the-art general relativistic magnetohydrodynamic (GRMHD) simulations of black hole accretion show that jets initially emerge with wide-opening angles near the central BH and are subsequently collimated and accelerated into relativistic narrow beams by converting the Poynting energy to the kinetic one, with the help of external pressure support from disk winds \citep[e.g.,][]{mckinney2006general, tchekhovskoy2008simulations}. Since such active collimation and acceleration processes are expected to take place at compact scales typically within $\lesssim 10^4-10^6$ Schwarzschild radii ($r_{\rm s}$) from the central BH \citep[e.g.,][]{marscher2008inner}, high-resolution observations using very long baseline interferometry (VLBI) are especially crucial to directly access the relevant sites. 

In recent years, VLBI studies of jet collimation have been performed for an increasing number of AGNs, including radio galaxies such as M\,87 \citep{junor1999, asada2012structure, hada2013innermost, nakamura2018parabolic,hada2024m}, NGC\,315 \citep{park2021jet, boccardi2021jet, ricci2022exploring}, NGC\,4261 \citep{nakahara2018finding, yan2023kinematics}, NGC\,6251 \citep{tseng2016structural}, Cyg A \citep{nakahara2019cygnus, boccardi2016}, NGC\,1052 \citep{baczko2022ambilateral, nakahara2019two}, as well as quasars such as 3C\,273 \citep{okino2022collimation}, the narrow-line Seyfert 1 galaxy 1H\,0323+342 \citep{hada2018collimation}, and a sample of blazars \citep{kovalev2020}. These studies commonly revealed that jets remain well-collimated (parabolic or cylindrical) near the BH but expand more rapidly at larger distances. Moreover, several sources provide evidence that their jet collimation and acceleration regions are cospatial, which is consistent with GRMHD predictions. 

Nevertheless, most sources studied so far belong to rather bright AGNs with powerful relativistic jets. The evolution of jet collimation as a function of accretion rate remains unclear. Expanding jet collimation studies to lower-power jets (or lower accretion rate systems) is crucial for achieving a more comprehensive understanding of jet formation across a broader range of SMBH activity.

The nearby giant elliptical galaxy M\,84, located in the Virgo Cluster \citep[$D=16.83$ Mpc;][]{tully2016cosmicflows}, exhibits bipolar radio jet structures expanding across $13$ kpc to the northern and southern directions \citep{laing1987rotation,laing2011deep}, down to subpc scales~\citep{ly2004, wang2022multifrequency}. The observed H$\alpha$ luminosity of $L_{({\rm H\alpha})}\sim 2 \times10^{39} \rm \, erg \, s^{-1}$ \citep{bower1997nuclear} indicates that the galaxy hosts a genuine low-luminosity AGN (LLAGN) \citep[e.g.,][]{ho1997search}. Its center resides an SMBH with an estimated mass in the range of $(4-26) \times 10^8\,M_\odot$ \citep{bower1997kinematics, maciejewski2001kinematics, walsh2010supermassive}. While the precise mass of the central SMBH in M\,84 remains subject to debate, this paper adopts the latest value reported by \cite{walsh2010supermassive} of $8.5 \times 10^8\,M_\odot$. The nuclear bolometric luminosity of $\sim 10^{41}$ \rm erg\,s$^{-1}$ \citep{ho1999spectral}, corresponding to an Eddington ratio of $\sim 5 \times 10^{-7} \, L_{\rm Edd}$, indicates very low activity of the nucleus. Thanks to its proximity and high BH mass, M\,84 is one of the very few objects in which Bondi radii of the central SMBH can be spatially resolved by X-ray observations~\citep{russell2013radiative, bambic2023agn}. Besides, VLBI observations provide angular resolution sufficient to probe gravitational scales close to the black hole (one milliarcsecond (mas) $\approx 0.082\,{\rm pc} \approx 10^3\,r_{\rm s}$). These features make M\,84 an exceptional target for studying jet formation regions of a very weakly accreting BH in detail. 

In this paper, we report the results from our multifrequency Very Long Baseline Array (VLBA) observations of M\,84 in order to explore the jet collimation property in a lower accretion regime. The contents of this paper are organized as follows: in Section\,\ref{sec:obs} we summarize the observation and dataset, in Section\,\ref{sec:results} we report the result of this work, in Section\,\ref{sec:discussion} we discuss our interpretation of the result in the astrophysical context, and we present our conclusions in Section\,\ref{sec:summary}. Throughout this paper, we adopt the following conventions for cosmological parameters: $H_0=70$ km s$^{-1}$, $\Omega_{\rm M}=0.25$, and $\Omega_\Lambda=0.75$ in a flat Universe, yielding a linear-to-angular scale conversion of 0.082 pc mas$^{-1}$. 

\section{Observations and Data Reduction} \label{sec:obs}
\subsection{Multi-frequency VLBA data} \label{sec:vlba}

On 2014 March 26, we observed M\,84 with VLBA during an observing program of M\,87 (hereafter referred to as VLBA I). M\,84 was observed as a phase-referencing source of M\,87 to measure the relative core shift between the two sources~\citep[e.g.,][]{hada2011origin}. In this paper, we regard M\,84 as a target and M\,87 as a phase-referencing calibrator. 

The observations were made at seven frequencies of 1.4\,GHz (L-band), 2.3/8.4\,GHz (S/X bands), 5\,GHz (C-band), 15\,GHz (U-band), 24\,GHz (K-band), and 43\,GHz (Q-band) in a 9 hr continuous run. The receiver at each frequency band was alternated in turn every 10–30 minutes. The observation involved rapid switching between M\,87 and M\,84 (separated by $1.5^{\circ}$ on the sky) to allow phase-referencing and relative astrometry between the two sources. Some short scans on the bright source OJ\,287 were also inserted as a fringe finder and bandpass calibrator. The observation was made in dual (left-/right-hand circular) polarization mode, and the received signals were sampled with 2-bit quantization and recorded at an aggregate rate of 2048 Mbps (a total bandwidth of 256 MHz per polarization). The down-converted signals were divided into two 128\,MHz sub-bands in each polarization. Unfortunately, the S-band data were severely affected by radio frequency interference, so we were forced to discard the S-band data in the subsequent analysis.

The initial data calibration was performed using the National Radio Astronomy Observatory (NRAO) Astronomical Image Processing System (AIPS; \citealt{greisen2003}) based on the standard VLBI data reduction procedures. The amplitude calibration was made with opacity corrections using the measured system noise temperature and the elevation-gain curve of each antenna. We then applied a priori corrections to the visibility phases, accounting for antenna parallactic angle differences between M\,84 and M\,87, correcting ionospheric dispersive delays using the Jet Propulsion Laboratory (JPL) ionospheric model, and addressing instrumental delays and phases based on calibration scans of OJ\,287. Then, to create images of these sources, we performed a fringe-fitting on each source separately and removed residual delays, rates, and phases assuming a point source model. {\tt CLEAN} images were created in the DIFMAP software \citep{shepherd1997difmap} with iterative phase/amplitude self-calibration.

\subsection{Atmosphere-corrected phase-referencing} \label{sec:atm}
Regarding relative astrometry between M\,84 and M\,87, our phase-referencing analysis at each frequency was performed in the following way. After the a priori corrections of the parallactic angle, ionosphere, and instrumental delay, we performed a fringe-fitting only on M\,87 with its source model created by the previous imaging process. With this procedure, we derived the solutions for residual phases/delays/rates, by which the effect of the M\,87 source structure was properly taken into account. These solutions were transferred to the time segments of M\,84. After that, we ran the AIPS task {\tt CALIB} on M\,87 (with the source model) to derive the corrections of time-dependent antenna gains for each antenna. The derived gain corrections were then transferred to the scans of M\,84. 

\begin{figure}
  \centering
  \includegraphics[width=\columnwidth]{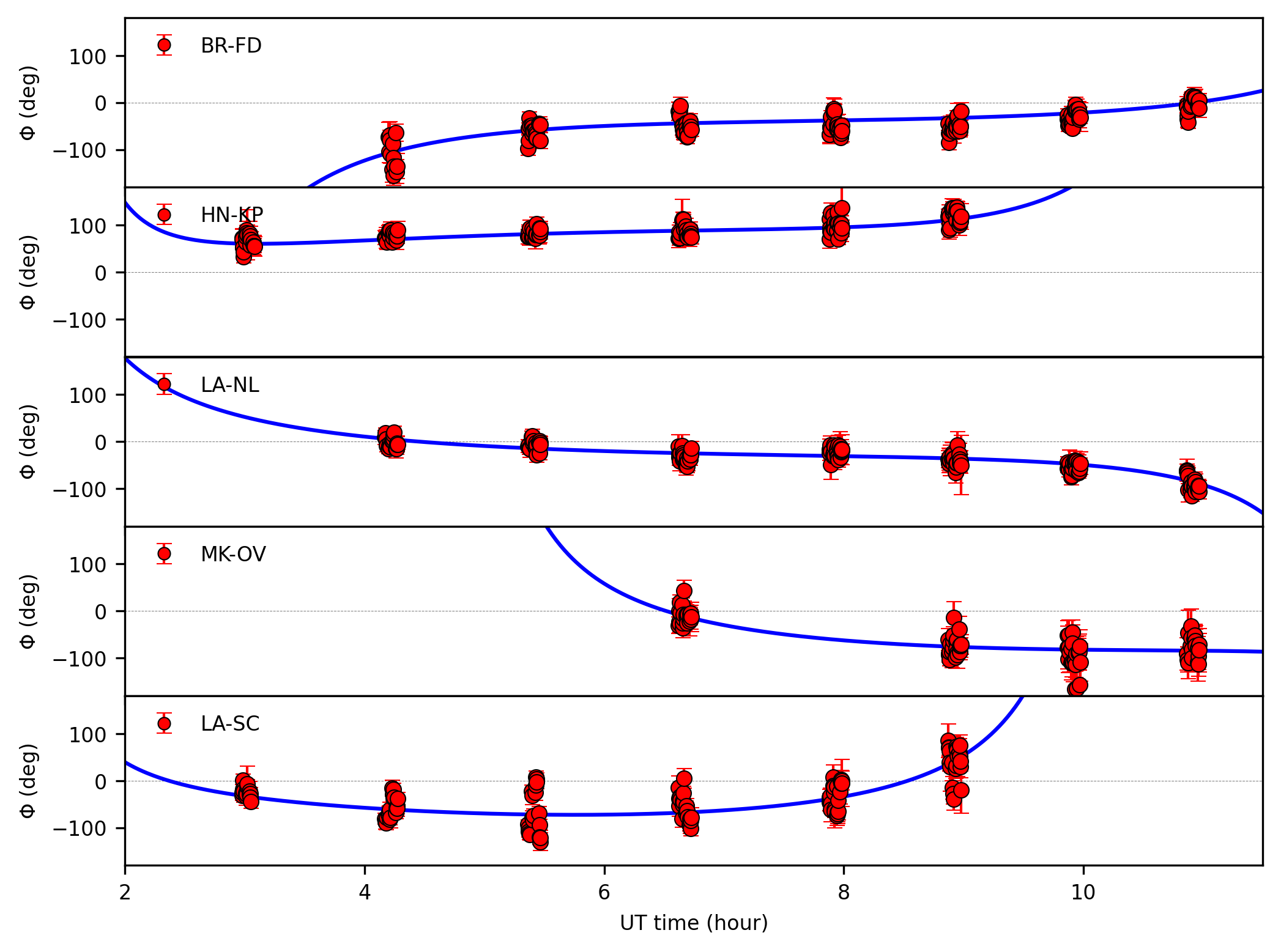}
  \includegraphics[width=\columnwidth]{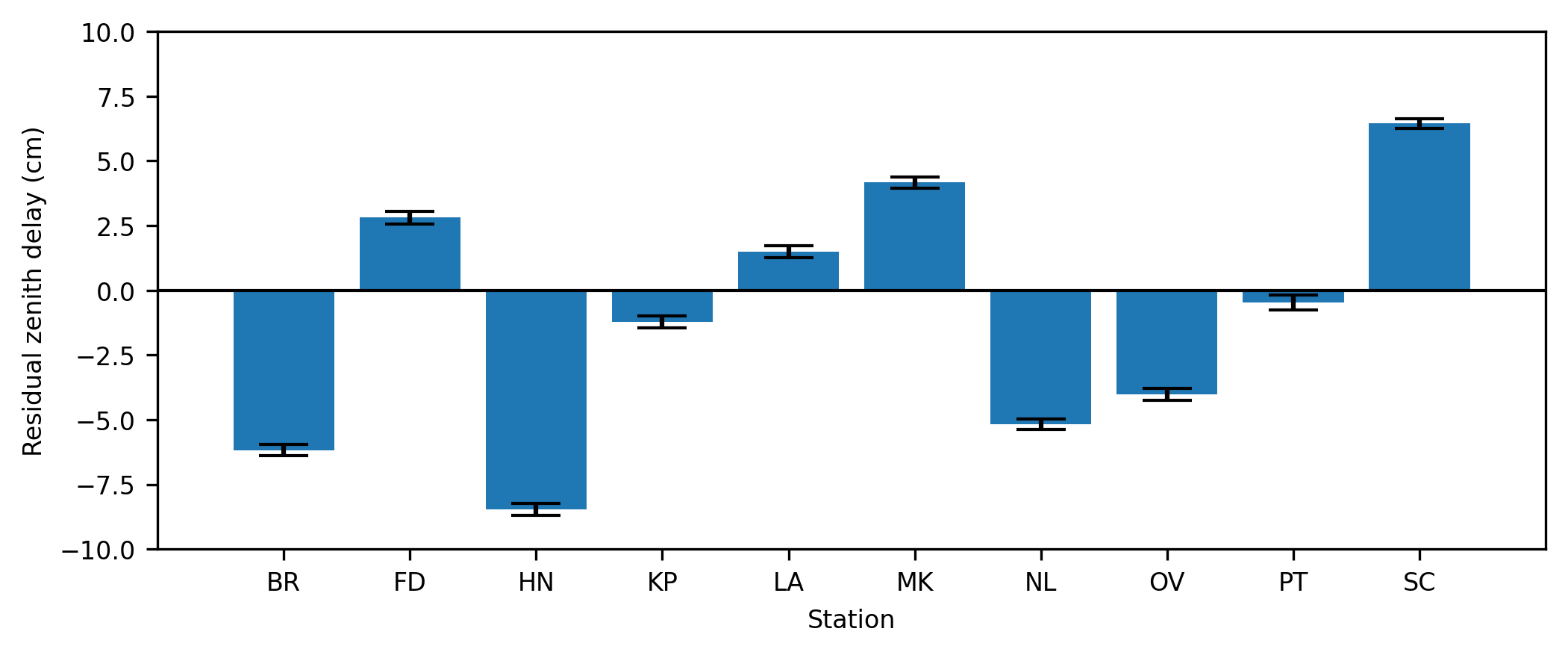}
  \caption{({\it Top}) Phase-referenced phases of M\,84 at 15\,GHz and fitted curves on five representative baselines. ({\it Bottom}) Zenith tropospheric delay value for each antenna derived from the phase-fitting method. The phase fitting was performed on all 45 VLBA baselines simultaneously.}
  \label{fig:atm_correct}
\end{figure}

\begin{figure*}
  \centering
  \includegraphics[width=16cm]{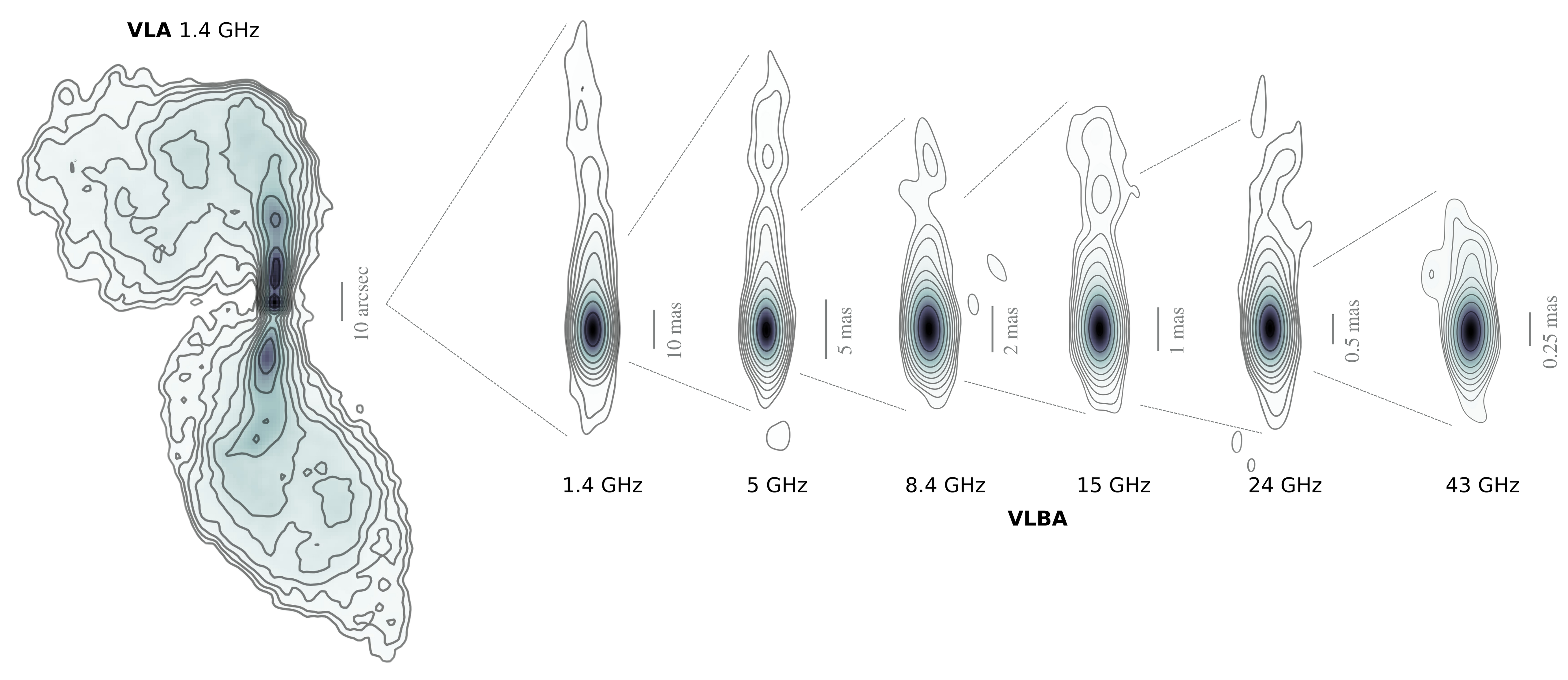}
  \caption{Multifrequency images of M\,84 from VLBA I and VLA observations. Contours are plotted at levels of (-1, 1, 2, 4, 8, 16, 32, 64, 128, 256) $\times$ 3$I_{\rm rms}$, where $I_{\rm rms}$ represents the rms noise level of the image listed in Table \ref{tab:paramsvlba} together with other image parameters.}
  \label{fig:fitsimages}
\end{figure*}

In this standard phase-referencing analysis, however, the quality of the phase-referenced images is still largely limited by inaccurate estimates of the zenith tropospheric delays ($\tau_{\rm 0, trop}$) in the correlator model (with a typical accuracy of the order of a few cm). In order to improve our relative position measurements, we further took additional calibration steps. We applied essentially the same procedure used in \citet{reid1999} and \citet{brunthaler2005}, where the differenced phase data between the target and calibrator were modeled. The phase-referenced phase of the target with respect to the calibrator is mainly a sum of the position offset and the effects of the zenith atmospheric delay error if the effects of the source structure are adequately modelled. While the phase induced by the position offset shows a simple daily sinusoidal pattern, the phase from the zenith delay error follows a more complex dependence on the zenith angle. Owing to the different behavior of the two contributions, it is possible to separate both effects and estimate a residual zenith delay error at each antenna by performing a least-square fitting of a model to the phase-referenced phase data. We developed a Python script to perform this fitting outside the AIPS by reading the phase-referenced phase data of M\,84. 

For a derivation of the residual tropospheric zenith delay error ($\Delta\tau_{\rm 0, trop}$) for each antenna from the above method, the use of single frequency data is adequate since the tropospheric delays are non-dispersive and our observations at six frequencies were made in a single observing track, so $\Delta\tau_{\rm 0, trop}$ should be common to all frequencies. Here we made use of the phase data at 15\,GHz since it is optimal in that (1) the phase signal-to-noise ratio (SNR) is sufficiently higher than those at 24/43\,GHz; (2) the observed coverage of zenith angle is sufficiently large; and (3) the contribution from the residual ionosphere component is sufficiently smaller than those at the lower frequencies. In Figure~\ref{fig:atm_correct}, we show phase-referenced phases of M\,84 at 15\,GHz on some example baselines and corresponding best-fit curves as well as the derived residual values of zenith tropospheric delay for each antenna. One can see that $\Delta\tau_{\rm 0, trop}$ varies from station to station, ranging from 0.2 cm to 7.5 cm. 

The derived $\Delta\tau_{\rm 0, trop}$ value for each antenna was then exported back to AIPS to correct the raw visibility data at all the observed frequencies. This correction was applied by using the AIPS task {\tt CLCOR} with {\tt OPCODE=`ATMO'}. With these further corrected visibility data, we again performed a phase-referencing analysis and produced phase-referenced images of M\,84 at each frequency. In Figure\,\ref{fig:PRSTR vs PRATM} we show a comparison of the standard phase-referencing and atmosphere-corrected phase-referencing images of M\,84 at each frequency. As clearly seen, the image dynamic range of the atmosphere-corrected phase-referencing images is greatly improved, which permits us to measure the core position at a higher accuracy. The image improvement is especially greater at higher frequencies since the total error was dominated by the tropospheric errors. On the other hand, the improvement of the lower-frequency images is relatively modest. This is because the total error budget of the low-frequency images is dominated by the residual dispersive ionospheric delay component rather than the troposphere. For details of core shift measurement using the phase-referenced images, see Section~\ref{ssec:coreshift}.

\subsection{VLA archival image} \label{sec:vla}
In addition, we utilized VLA archival data from NASA/IPAC Extragalactic Database. The observation was performed on November 9, 1980, in 1.4\,GHz \citep[][]{laing1987rotation}.  The complete multifrequency images from the VLBA I and VLA observations are shown in Figure \ref{fig:fitsimages}.

\begin{table}[h]
    \small
    \begin{threeparttable}
    \caption{Multifrequency Radio Observations of M84\\}
    \begin{tabularx}{\columnwidth}{c c c c c}
    \hline
    \hline
    Frequency & Band & Beam Size & \, $I_{\rm p}$ & $\quad I_{\rm rms}$ 
    \\
    (GHz) & & (mas $\times$ mas) & \, ($\mathrm{\frac{mJy}{beam}}$) & \quad ($\mathrm{\frac{mJy}{beam}}$) 
    \\
     & & (a) & \, (b) & \quad(c)
    \\ \hline
    \multicolumn{5}{l}{\; \textbf{VLBA I}} 
    \\ \hline
    1.4 & L & $10\times5$ & \, 80 & \quad 0.23    
    \\
    5 & C & $3\times1.5$ & \, 110 & \quad 0.14
    \\
    8.4 & X & $2 \times 1$ & \, 104 & \quad 0.65
    \\
    15 & U & $1\times0.5$ & \, 105 & \quad 0.34
    \\
    24 & K & $0.7 \times 0.35$ &\, 109 & \quad 0.41
    \\
    43 & Q & $0.34 \times 0.17$ & \, 94 & \quad 0.76
    \\\hline
    \multicolumn{5}{l}{\; \textbf{VLA}} 
    \\ \hline
    1.4 & & $3860 \times 3860$ & \, 152 & \quad 1.20
    \\ \hline
    \end{tabularx} \label{tab:paramsvlba}
    \tablecomments{(a) Synthesized beam with naturally weighted scheme. (b) Peak intensity of the self-calibrated images of M84 under the naturally weighted scheme. (c) The rms image noise level of M84 images under the naturally weighted scheme.}
    \end{threeparttable}
\end{table}

\subsection{Supplementary VLBA data}
We also utilized supplementary VLBA observations previously published by \cite{wang2022multifrequency} (hereafter referred to as VLBA II). These observations consist of simultaneous multifrequency VLBI data obtained on June 2, 2020, at 5 and 22\,GHz, and on March 31, 2021, at 44 and 88\,GHz. The inclusion of the 88\,GHz observations was intended to maximize the frequency coverage for improved analysis across a broad range. For more details of data analysis refer to \cite{wang2022multifrequency}.

\section{Image analysis and results} \label{sec:results}
\subsection{Jet collimation profile}\label{ssec:collimation}

\begin{table*}
    \small
    \begin{threeparttable}
    \caption{Jet Collimation Profile Fitting Parameters\\}
    \begin{tabularx}{\textwidth}{P{3cm}*{8}{Q{1.75cm}}}
    \hline
    \hline
    & $A_0$ & $a$ or $a_{\rm u}$  & $a_{\rm d}$ & $r_0$ & $W_0$ & $n$ & $\chi^2/{\rm dof}$
    \\ \hline
    \textbf{Single power-law}  & $0.09 \pm 0.01$ & $1.12 \pm 0.01$ & ... & ... & ... & ... & 2.59
    \\ \hline 
    \textbf{Broken power-law}& ... & $0.71\pm{0.03}$ & $1.16\pm{0.01}$ & $15.8\pm{3.0}$ & $1.45\pm0.28$ & 75 & 1.09   
    \\\hline \hline
    \end{tabularx}
    \tablecomments{$A_0$: normalization parameter, $a$: power-law index of the single power-law model, $a_{\rm u}$: upstream power-law index of the broken power-law model, $a_{\rm d}$: downstream power-law index of the broken power-law model, $r_0$: de-projected distance of the break position, $W_0$: jet radius at the break position, $n$: controlling parameter for the sharpness of the break, and $\chi^2/{\rm dof}$: reduced chi-square of the fitting where ${\rm dof}$ is the degree of freedom.}
    \end{threeparttable}
    \label{tab:fitting_result}
\end{table*}
We investigated the M\,84 jet collimation by deriving the jet radius as a function of position along the NS direction as follows. Using the AIPS task {\tt SLICE} we sampled the pixel-based slice profile of the jet. A Gaussian function was fitted to the profile to derive the resultant fitted Gaussian. Since the map is convolved by the beam size corresponding to the angular resolution, the jet radius (half of the jet width) is thus defined as half of the FWHM of the resultant fitted Gaussian, $\Phi_0$, deconvolved from the beam FWHM, $\Phi_{\rm b}$, can be written as $W_{\rm j}=\sqrt{\Phi_0^2 - \Phi_{\rm b}^2}$. The uncertainty is estimated from the combination of the Gaussian fitting error and the imaging error that is estimated as $\Phi_{\rm b}/5$. The jet radii are plotted with respect to the location of the central engine by assuming de-projected distance due to the inclination angle $i=74^{\circ}$ \citep{meyer2018detection}.

\begin{figure*}
  \includegraphics[width=\textwidth]{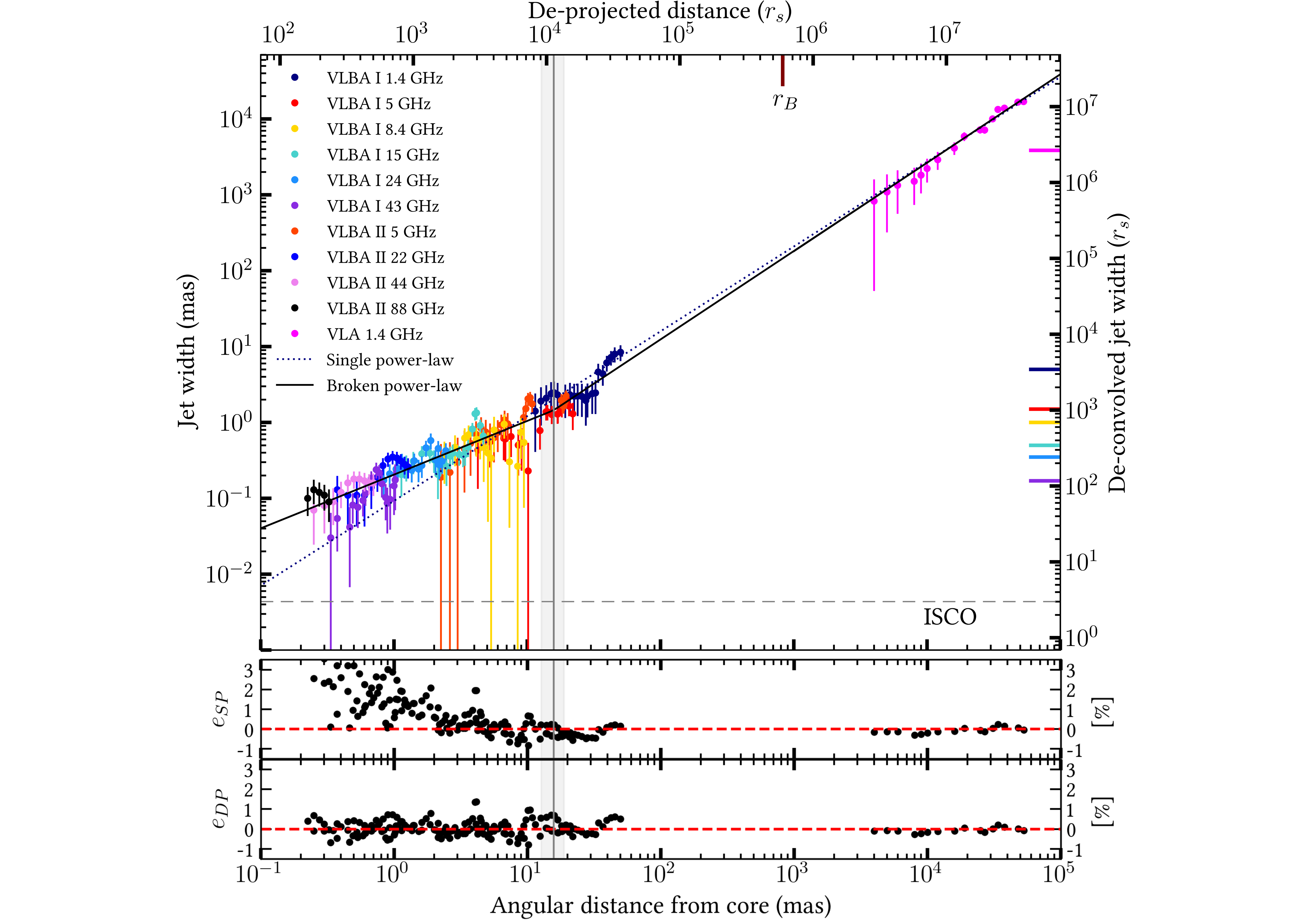}
  \caption{The upper panel shows the jet width of M\,84 jets as a function of de-projected distances from the black hole in units of mas and $r_{\rm s}$. The solid black line represents the best fit of the double power-law model while the dotted line shows the best fit of the single power-law model. In the double power-law fitting, the jet transitions from a semi-parabolic profile ($W \propto r^{0.71}$) in the upstream to a conical profile ($W \propto r^{1.16}$) in the downstream. The vertical grey line indicates the transition occurring at $\sim 1.67 \times 10^4\,r_{\rm s}$, with the grey shaded area denoting the associated error. The red tick at the top axis marks the Bondi radius $r_{\rm B}$ of M\,84 at $\sim 5.9 \times 10^5\,r_{\rm s}$. The dashed horizontal line indicates the location of the innermost stable circular orbit (ISCO) for a non-rotating black hole. The middle and lower panels depict the residuals of the single power-law fit and double power-law fit, respectively.}
 \label{fig:jetcollimation}
\end{figure*}

In Figure \ref{fig:jetcollimation}, we demonstrate how the jet collimation profile evinces the evolution of the global jet structure. We performed regression analysis with single and double power-law models throughout the whole VLBI data set and showed the results and their relative residuals. The jet width plot relatives to de-projected distance from the black hole can be easily fitted with a single power-law model:
\begin{equation}
    W_{\rm j}(r)=A_0 \, r^{a},
\end{equation}
where $A_0$ is a constant and $a$ is the power-law index. $A_0$ and $a$ are free parameters. The single power-law function gives the best-fit by a power-law index of $a=1.12 \pm 0.01$ and $A_0=0.09 \pm 0.01$ with reduced Chi-squared value of $\chi^2 / {\rm dof}=2.59$ where ${\rm dof}$ is the degree of freedom. This indicates that the single power-law model does not adequately explain the data.

Next, we applied a double power-law model in the form of a broken power-law function from \cite{tseng2016structural} with a little modification as:
\begin{equation}
      W_{\rm j}(r)=W_0\left[\left(\frac{r}{r_0}\right)^{a_{\rm u} n}+\left(\frac{r}{r_0}\right)^{a_{\rm d} n}\right]^{1/n},
\end{equation}
where $a_{\rm u}$ is the upstream power-law index, $a_{\rm d}$ is the downstream power-law index, $r_0$ is the break position, $W_0$ is the jet radius at the break position, and $n$ is the sharpness parameter. The best-fit parameters of the power-law models yield a transition from a semi-parabolic, $a_{\rm u}=0.71\pm{0.03}$, to a conical function, $a_{\rm d}=1.16\pm{0.01}$, at the distance of $r_0=15.8\pm{3.0}$ mas $\approx \,1.67 \times 10^4 \, r_{\rm s}$. The jet width at the transition point is $W_0=1.45\pm0.28$ mas $\approx \, 1.46\times10^3 \, r_{\rm s}$ with the sharpness of the break of $n=75$. The broken power-law improved the residuals and reduced Chi-squared for the data to $\chi^2/{\rm dof} =$ 1.09. Despite many years of difference, the data set shows consistency and hence constrains the result.

\subsection{Core shift}\label{ssec:coreshift}
\begin{figure}
  \centering
  \includegraphics[width=8.5cm]{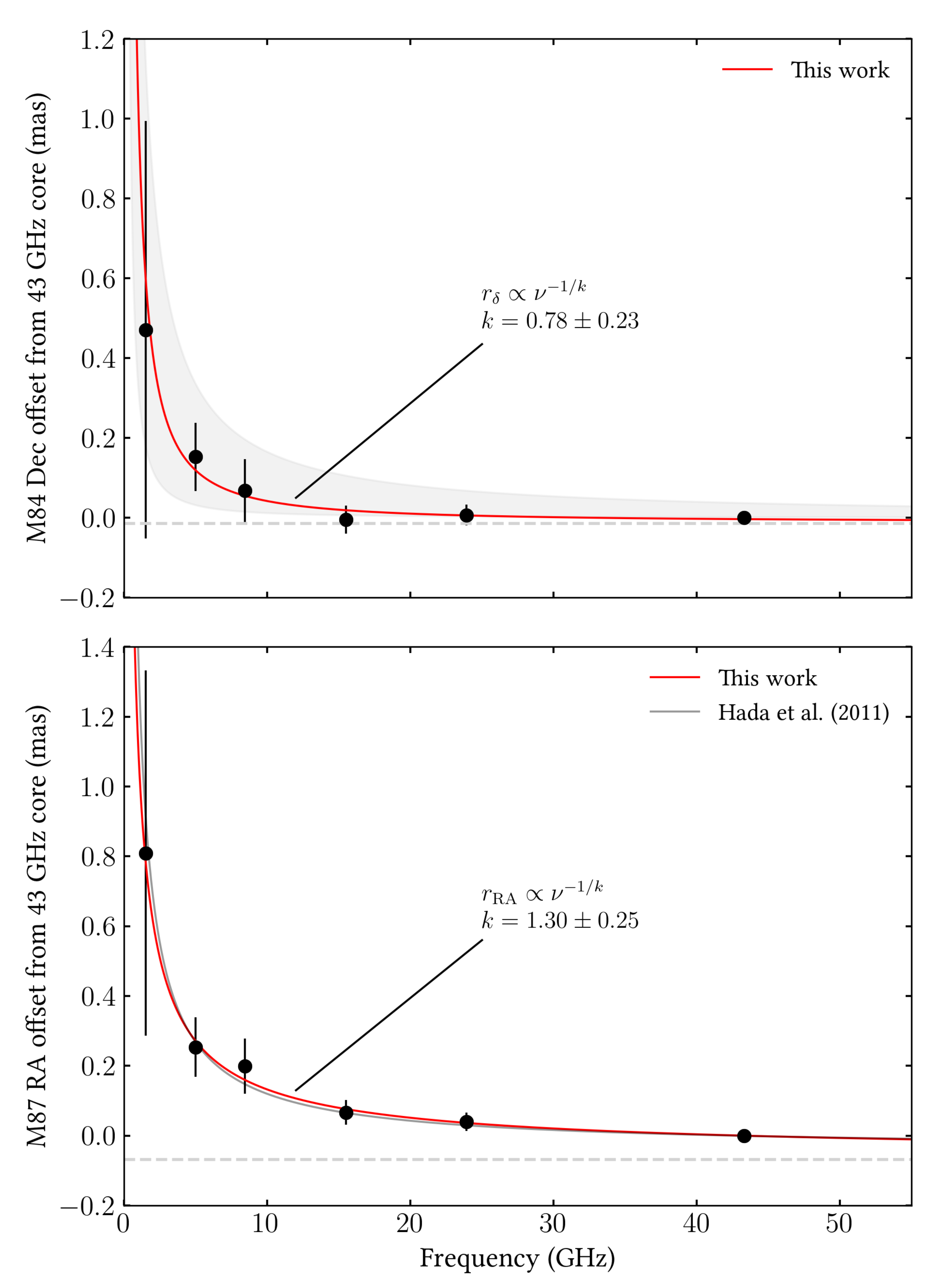}
  \caption{({\it Top}) Core position of M\,84 as a function of frequency in declination direction. The shaded grey area represents uncertainties arising from the M\,87 jet PA$\,\approx 270^{\circ}-320^{\circ}$. ({\it Bottom}) Core position of M\,87 as a function of frequency in right ascension direction compared to the previous work by \cite{hada2011origin}. The red lines of the fitted result are based on the assumption that the M\,87 jet PA$\,=290^{\circ}$. The dashed lines denote the asymptotes of the solid curves as they converge toward the black hole location.}
  \label{fig:coreshift}
\end{figure}

To construct a precise jet collimation profile, it is crucial to accurately determine the true location of the central engine. The position of the radio core can be identified as the point at which the optical depth due to synchrotron self-absorption reaches unity \citep{blandford1979relativistic}. This core is the manifestation of the true location of the central engine mixed with the multi-component jet base. If we assume a conical jet and the central engine is located at the jet apex, the location of the radio core follows  $r_{\rm c}(\nu) \propto \nu^{-1/k_{\rm r}}$ \citep{lobanov1998ultracompact, o2009magnetic}. Thus, the location of the radio core is frequency-dependent and can be specified through multi-frequency observations. Due to the weak nature of our source and the absence of bright jet components, self-referencing technique employed in some earlier studies \citep[e.g.,][]{croke2008aligning, o2009three} is not applicable. Thus, phase-referencing is essential for the core-shift measurement of M\,84.

We measured the core position using the visibility domain modelfit function on Difmap, adopting a single elliptical Gaussian model. Subsequently, we compared the result with an image domain elliptical Gaussian fitting via {\tt JMFIT} in AIPS. Both methods yielded consistent results within the error bar scale. It should be noted that the core represents the peak flux density, which includes combined emissions from the jets and the central engine. To determine the true position of the core, phase-referencing images were utilized, taking into account the core shift effect resulting from both the target, M\,84, and reference, M\,87, sources. We assumed that the M\,84 jet has solely a declination component (jet position angle $(\mathrm{PA}) = 0^\circ$), while the M\,87 jet possesses effects in both RA and declination directions \citep[$\mathrm{PA} = 290^\circ$;][]{hada2013innermost}. Therefore, we were able to decouple the core shift effect from both objects.

\begin{table}[h]
    \small
    \centering
    \caption{Total Error Budgets\\}
    \begin{tabular}{P{1.9cm}*{6}{Q{0.7cm}}}
    \hline
    \hline
    & \multicolumn{6}{c}{Positional uncertainties ($\mu$as)}
    \\ \cline{2-7}
      & 1.4 GHz & 5.0 GHz & 8.4 GHz & 15 GHz & 24 GHz & 43 GHz
    \\ \hline
     Beamwidth/ SNR & 437 & 77 & 72 & 31 & 21 & 14
    \\ \hline
    Ionospheric residuals & 272 & 25 & 9 & 3 & 1 & 0.3     
    \\ \hline
    Tropospheric residuals & 9 & 9 & 9 & 9 & 9 & 9 
    \\ \hline
    Core identification & 93 & 25 & 15 & 14 & 12 & 6
    \\ \hline
    Earth orientation & 1 & 1 & 1 & 1 & 1 & 1
    \\ \hline
    Antenna position & 2 & 2 & 2 & 2 & 2 & 2 
    \\ \hline
    Apriori source coordinates & 2 & 2 & 2 & 2 & 2 & 2
    \\\hline
    \textbf{Root sum squared} & 523 & 85 & 75 & 35 & 26 & 18
    \\ \hline
    \end{tabular}
    \label{tab:total error}
\end{table}

The result of core shift measurements is shown in Figure \ref{fig:coreshift}. The core shift effect of M\,84 can be described by a power-law fitting of $r_{\delta}(\nu)=A\nu^{-1/k}+B$ mas, resulting in best-fit parameters of $k=0.78 \pm 0.23$, $A=1.03 \pm 0.53$, and $B=-0.012 \pm 0.014$. The error bar for each data point is determined based on various effects outlined in Table \ref{tab:total error}. The separation between the M\,84 43\,GHz core and 1.5\,GHz core falls within the range of $\lesssim 1$ mas. Simultaneously, we obtained the core shift effect of M\,87. The result shows good consistency with a power-law index of $k=1.30 \pm 0.25$, $A=1.16 \pm 0.23$, and $B=-0.063 \pm 0.028$ for the right ascension component. The comparison with the previous core shift measurement of \cite{hada2011origin} demonstrates good consistency within the error bar of the two measurements.

Based on this core shift fitting result, the separation between the central SMBH and the 43\,GHz core provides a clue about the true location of the central engine. Although still a subject of debate, estimates for the viewing angle of M\,84 range from $58^\circ$ to $74^\circ$ \footnote{We adopt a jet viewing angle of $i=74^{\circ}$ for the collimation profile. Using $i=58^{\circ}$ instead would shift de-projected distances by $\sim10\%$, without affecting our overall conclusions.} \citep{laing2014systematic,meyer2018detection,wang2022multifrequency}, giving us an upper limit for the BH location at $\lesssim 0.01$ pc ($\lesssim 14 \, r_{\rm s}$) from the highest observational frequency. This value is comparable or even smaller than the BH location distance from the 43\,GHz core of M\,87, where the location is believed to be around $0.007-0.01$ pc \citep[$14-23 \, r_{\rm s}$;][]{hada2011origin}.

\section{Discussion} \label{sec:discussion}
\subsection{On M\,84 jet profile} 
\label{ssec:jetprofile_discuss}

\begin{figure}[ttt]
  \centering
  \includegraphics[width=8.5cm]{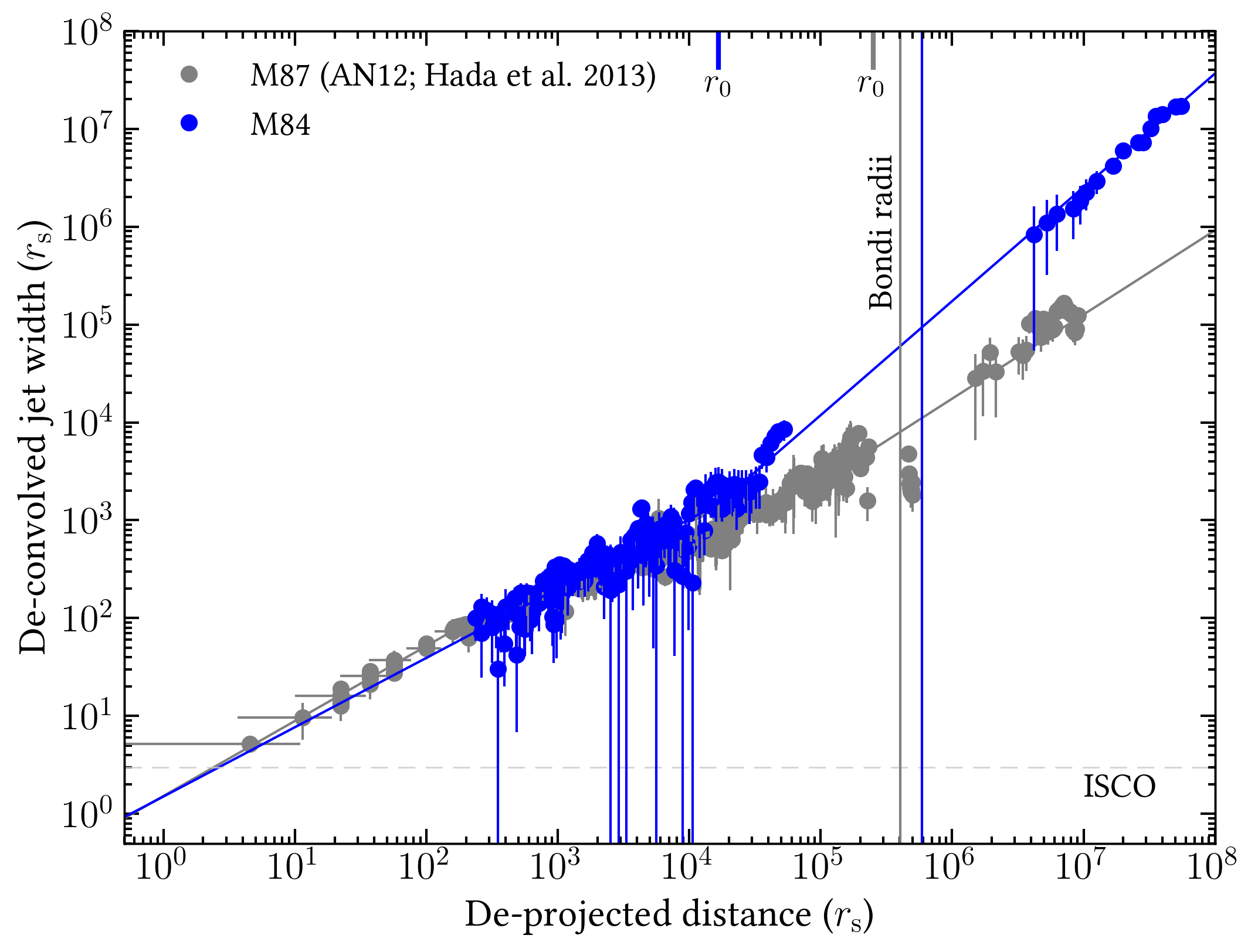}
  \caption{Comparison between the jet collimation profiles of M\,84 and M\,87. The blue and grey ticks at the top axis mark the transition radius of M\,84 and M\,87. The blue and grey vertical line marks the location of Bondi radii of M\,84 and M\,87 \citep[$r_{\rm B}\sim4.0\times10^5\,r_{\rm s}$;][]{allen2006relation}, respectively. The jet collimation profile of M\,87 shows a transition in the same order as its Bondi radius, whereas in M\,84 the transition occurs at a significantly smaller distance before reaching its Bondi radius.}
  \label{fig:m84vsm87}
\end{figure}

In recent years, the collimation profile of relativistic jets has extensively been investigated for a number of AGN sources. In particular, a morphological transition of the jet from parabolic to conical shape at certain distances from the jet base has been discovered in a growing number of jet sources with various AGN types (see references in Section\,\ref{sec:intro}). This suggests that such a morphological transition is likely a common feature of AGN jets regardless of AGN type. 

In the present study, we found a similar jet transition also in M\,84, which is at the weakest end among the sources where the jet collimation profile has been measured so far. The formation of a parabolic jet is in agreement with recent GRMHD simulations of jet production from radiatively-inefficient-accretion flows~\citep[RIAF; e.g.,][]{pu2022, nakamura2018parabolic}. In MHD jets, such a collimation zone is physically linked to the efficient bulk acceleration of the jets. For M\,84, in fact, a recent study of jet kinematics by \citet{wang2022multifrequency} indicates a hint of gradual acceleration of jet components within several mas from the core, where the jet shape is parabolic. Therefore, the M\,84 jets may have a cospatial collimation and acceleration zone as suggested by the theories. The co-existence of jet collimation and acceleration is also indicated in other jet sources (e.g., M\,87 \citep{nakamura2013parabolic, asada2013discovery, mertens2016}; NGC\,315 \citep{park2021jet}; 1H\,0323+342 \citep{hada2018collimation}). 

In contrast, the parabolic profile of the M\,84 jet breaks around $\sim$10$^4$\,$r_{\rm s}$ or $\sim$1\,pc from the jet base, much shorter than those seen in other sources. For comparison, in Figure~\ref{fig:m84vsm87} we overlay the observed M\,84 jet collimation profile on that of another FR-I radio galaxy M\,87, whose jet geometry is accurately known. While the M\,87 jet break occurs at $\sim 2.5 \times 10^5\,r_{\rm s}$ \citep[45 pc;][]{asada2012structure} from the jet base, one can clearly see that the M\,84 jet breaks almost two orders of magnitude shorter distances than M\,87. According to the standard paradigm of MHD jets, the shape of the jet is determined by the balance between the jet internal pressure and the ambient pressure from the external medium \citep[e.g.,][]{komissarov2009magnetic}. One of the hotly debated scenarios is that the gas bounded by the gravitational potential of SMBH supports the parabolic jet, which triggers a jet transition around the Bondi radius or the sphere of gravitational influence (SGI). Then here we first discuss the relationship between the observed jet break distance and Bondi/SGI radii of M\,84. 

In order to estimate the area in which the gravitational potential of the central engine becomes dominant, one can write
\begin{equation}
    r_{\rm SGI}=\frac{GM}{\sigma^2}, 
\end{equation}
where the $\sigma$ is the velocity dispersion of stars around SMBH. Under the assumption that the stellar distribution is supported by random motion with the velocity dispersion of the collective stars in M\,84 of $\sigma \simeq 296 \pm 14$ km s$^{-1}$ \citep{walsh2010supermassive}, we obtained $r_{\rm SGI}=5.13 \times 10^5\,r_{\rm s}$ as an upper limit.

If we simply assume the spherical accretion radius where the core becomes thermodynamically stable and thus virial equilibrium is satisfied, the Bondi radius can be derived by
\begin{equation}
    r_{\rm B}=\frac{2GM_{\rm BH}}{c^2_{\rm s}}, 
\end{equation}
where $G$, $c_{\rm s}=\sqrt{\gamma_1 k T/ \mu m_{\rm p}}$, $T$, $\mu$, $m_{\rm p}$, and $\gamma_1$ are the gravitational constant, the adiabatic sound speed, the gas temperature, the mean atomic weight, the proton mass, and the adiabatic index of the gas, respectively. From the newest data in 2019 based on the measurement of the north, south, west, and east parts of the galaxy, the inferred value of M\,84 Bondi radius is $r_{\rm B} \simeq 5.9 \times 10^5\,r_{\rm s}$ with Bondi accretion rate of $\dot{M}_{\rm B}= 3.74^{+1.05}_{-0.89} \times 10^{-3} M_{\odot} \, \rm yr^{-1}$ \citep{bambic2023agn}. This measurement provides tighter error bars and in agreement with \cite{allen2006relation} ($\dot{M}_{\rm B}= 8.5^{+8.4}_{-4.1} \times 10^{-3} M_{\odot} \, \rm yr^{-1}$) and recent study by \cite{plvsek2022relation} ($\dot{M}_{\rm B}= 2.4^{+1.9}_{-1.5} \times 10^{-3} M_{\odot} \, \rm yr^{-1}$).

By these estimations, the transition in M\,84 occurs well before the jet reaches $r_{\rm SGI}$ and $r_{\rm B}$. This challenges the scenario that a jet collimation break occurs near the Bondi radius, suggesting that the situation is not so simple. Recently, a similar jet transition sufficiently below the Bondi radius has also been reported in other LLAGN NGC\,4261~\citep{yan2023kinematics} and NGC\,1052~\citep{baczko2022ambilateral,baczko2024putative}. We will discuss this in more detail in Section~\ref{ssec:comparison}. 

\subsection{Comparison with other LLAGN sources} \label{ssec:comparison}

\begin{figure}
  \centering
  \includegraphics[width=8.5cm]{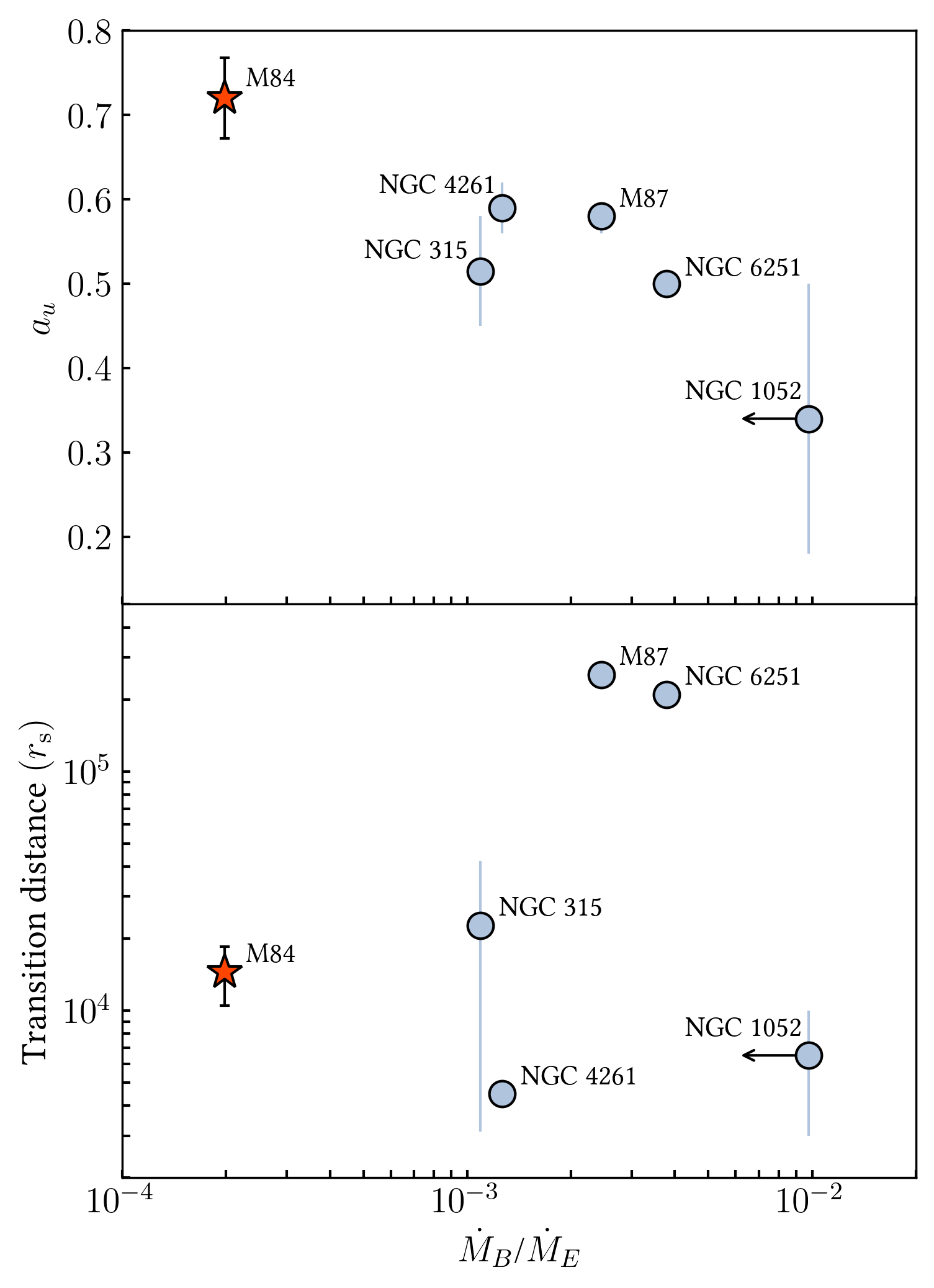}
  \caption{({\it Top}) Upstream power-law indices to normalized accretion rate among various objects in the LLAGN class. ({\it Bottom}) Transition break locations to normalized accretion rate among various objects in the LLAGN class. The black-capped error bar on the M\,84 data point denotes our measurement uncertainty. Values for NGC\,315, NGC\,4261, and NGC\,1052 represent averages from multiple studies, and the grey vertical line represents data point discrepancies. Transition distances adjusted based on the latest BH masses: NGC\,4261 \citep{sawada2022circumnuclear}; NGC\,315 \citep{boizelle2021black}; M\,87 \citep{akiyama2019first}. All parameters of the LLAGN objects obtained from the literature are listed in Table~\ref{tab:LLAGNtable}.} 
  \label{fig:LLAGNs}
\end{figure}

Our study on M\,84 has extended the parameter space of AGN jet collimation down to the lower accretion and core luminosity regime. This allows us to discuss whether there is any correlation between the characteristics of jet collimation and the central AGN activity in different power ranges.  

In Figure \ref{fig:LLAGNs}, we compiled the observed power-law indices of the parabolic jet and the transition distance from the central BH for various AGNs as a function of normalized accretion rates. The object parameters are provided in Table Appendix II.1. Here we limit our comparison only to LLAGNs (which meet $L_{\rm (H\alpha)}\lesssim 10^{40} \rm \, erg \, s^{-1}$, corresponding to sufficiently sub-Eddington sources) since other AGN classes such as quasars and NLSy1s have different modes of accretion flows, which makes our comparisons much more complicated. Interestingly, one can see the apparent trend indicating that lower accretion sources tend to have shorter jet transition distances and larger jet power-law indices, with M\,84 representing the upper and lower ends of the plots. The plot illustrates M84 as a low-end jet source in terms of its accretion rate, further distinguishing it from other known sources. A tentative negative correlation is found between the upstream power-law index and the normalized accretion rate when including M\,84 ($\tau=-0.60$, $p$-value$=0.23$), which lies at the low end of the accretion rate. This highlights the importance of including low-accreting sources such as M\,84 in studies of jet geometry. In contrast, the correlation between the transition distance and the normalized accretion rate remains statistically insignificant ($\tau=0.40$, $p$-value$=0.48$), regardless of whether M\,84 is included. Note that NGC\,4261 may be rather peculiar/exceptional in our sample since it contains clear nuclear accretion disk features extending from $\sim$300\,pc down to $\sim$0.2\,pc scales \citep{ferrarese1996evidence, jones2000, haga2015determination}. NGC\,1052 is another exceptional source, characterized by a particularly dense environment, with an obscuring torus detected at $\sim 0.5$ pc \citep{kameno2001dense, kameno2003dense}. Its collimation profile and accretion rate are not yet well-determined, therefore we exclude NGC\,1052 in the statistical test. Such a distinctive dense environment in the aforementioned objects may add further external support to confine the inner jet, resulting in a smaller jet power-law index than that of M\,84. It is also worth noting that the normalized accretion values used in this paper are based on observations far from the BH (at the order of Bondi radius), with an efficiency of $\eta=0.1$. Since accretion rates are generally expected to decrease as one approaches the event horizon \citep[e.g.,][]{blandford1999fate}, we anticipate a similar trend for other LLAGNs if their accretion rate at the horizon scale were measurable. 

According to theory and simulations, winds play a crucial role in confining jets. In LLAGNs, confinement is attributed to external pressure provided by the disk wind from RIAFs. \cite{yuan2015numerical} performed 3D GRMHD simulations of hot accretion flows around a Schwarzschild BH and found that the maximum flux of wind is very significant, with a radial profile described by $\dot{M}_{\rm wind} \approx \dot{M}_{\rm BH}(r /20 r_{\rm s})$, for region where the gravitational force of the BH dominates \citep{bu2016hydrodynamical}. Referring to Figure \ref{fig:LLAGNs}, LLAGNs with smaller $\dot{M}$ are expected to exhibit lower wind production rates, leading to ineffective collimation with larger $a_{\rm u}$. Our study provides important clues that the normalized accretion rate may play a role in controlling jet collimation. Further investigations of LLAGN jet collimation profiles for other sources are needed to better clarify the trend. 

\subsection{Magnetic-field strength of M\,84 radio core} \label{ssec:coreshift_discuss}
One of the intriguing results of our study is that the observed apparent core shift of M\,84 appears to be rather small compared to M\,87, despite the larger jet viewing angle of M\,84. A core shift is naturally expected if the radio core observed at each radio frequency represents an optically thick surface of the synchrotron self-absorption in the inner jet, while the magnitude of core shift depends on various jet internal physical quantities (e.g., magnetic-field strength and electron number density) and geometrical information about the jet (viewing angle, opening angle, and distance to the source). Nevertheless, under a reasonable assumption, an observed core shift may give us a useful insight into the strength of magnetic field \citep{lobanov1998ultracompact, hirotani2005kinetic,o2009magnetic}.

For simplicity, here we consider the situation of energy equipartition between the magnetic field and particles in the radio core regions. In this case, an equipartition magnetic field strength at 1 pc of actual distance from the jet vertex can be calculated as: 

\begin{equation}
    B_1 \simeq 0.025 \left( \frac{\Omega_{r \nu}^3 (1+z)^2}{\delta^2 \, \varphi_{\rm obs} \sin^2 \theta} \right)^{\frac{1}{4}}, 
\end{equation}
where $\delta$, $z$, $\theta$, $\varphi_{\rm obs}$ are the Doppler factor, redshift, viewing angle, and the jet opening angle ($\varphi_{\rm obs}=\varphi \sin{\theta}$), respectively. $\Omega_{r \nu}$ is so-called the core position offset measure \citep{lobanov1998ultracompact}: 

\begin{equation}
    \Omega_{r \nu} = 4.85 \times 10^{-9} \frac{\Delta r_{\rm core} D_{\rm L}}{(1+z)^2} \left( \frac{\nu_1^{1/k_{\rm r}}\nu_2^{1/k_{\rm r}}}{\nu_2^{1/k_{\rm r}}-\nu_1^{1/k_{\rm r}}}\right) \mathrm{pc \, GHz}, 
\end{equation}
where $\Delta r_{\rm core}$ is the core shift between frequencies $\nu_2$ and $\nu_1$, $D_{\rm L}$ is the luminosity distance in pc, and $k_{\rm r}=1$ for equipartition. 

Adopting that $\delta \sim 1$~\citep{wang2022multifrequency}, $z = 0.0039$, $\theta = 74^{\circ}$ \citep{meyer2018detection}, and $\varphi_{\rm obs} \approx 4.6^{\circ}$ from our jet collimation analysis, $B_1$ for M\,84 is estimated to be $\sim 6.5$ mG at 1 pc. 

Our estimate described above appears to be consistent with a recent study by \cite{wang2022multifrequency} based on the synchrotron self-absorption theory which assumed a uniform magnetic field and an isotropic power-law distribution of electrons. The estimated turnover frequency of $\nu_{\rm m}= 4.2 \pm 0.2$\,GHz corresponds to a magnetic field strength of the order of $1 \sim 10$ mG and an electron density of $\sim 10^{5}$ cm$^3$.

Using the estimated magnetic field strength, we can further consider the properties of the jet launching region. Measuring BH magnetic flux $(\Phi_{\rm BH})$ directly is extremely challenging, however, we can simply assume $\Phi_{\rm BH} \approx \Phi_{\rm jet}$, through flux freezing approximation. Following \cite{chamani2021testing}, the poloidal magnetic flux threading parsec-scale jets, which are produced via the Blandford-Znajek mechanism, can be expressed as:
\begin{equation}
\Phi_{\rm jet}=8 \times 10^{33} f(a) [1+ \sigma]^{1/2}\left( \frac{M_{\rm BH}}{10^9 M_{\odot}}\right) \left( \frac{B}{\rm G} \right) \mathrm{\ G \ cm^{2}},
\end{equation}
where
\begin{equation}
f(a)= \frac{1}{a} \frac{r_{\rm H}}{r_{\rm g}} = \frac{1+(1-a^2)^{1/2}}{a}.
\end{equation}
The jet magnetization parameter, $\sigma$, represents the ratio of the Poynting flux to the kinetic energy flux and is given by $\sigma = (\Gamma \theta_{\rm j} /s)^2$. The black hole event horizon radius, $r_{\rm H}$, can be expressed as $r_{\rm H}=r_{\rm g}(1+(1-a^2_*)^{1/2})$ where $r_{\rm g}$ is the gravitational radius and $a_*$ is the spin parameter of the black hole.
Assuming $s=1$ in a fully spinning black hole, we find the magnetic flux to be $\Phi_{\rm jet} \sim 4.5 \times 10^{31}$ G cm$^{2}$. 

Magnetically arrested accretion (MAD) state can be realized if $\Phi_{\rm BH,MAD}=50(\dot{M}_{\rm BH} r_{\rm g}^2c)^{1/2}$ \citep{narayan2003magnetically, tchekhovskoy2011efficient}. Adopting the predicted accretion rate at the ISCO of $\dot{M}_{\rm BH}\approx8.5 \times 10^{-6} M_{\odot} \, \rm yr^{-1}$ \citep{bambic2023agn}, we obtain $\Phi_{\rm BH, MAD} \sim 2.5 \times 10^{31}$ G cm$^{2}$. This implies that despite M\,84 being at the low end of LLAGNs, the MAD state could still be achieved, highlighting the crucial role of a well-ordered magnetic field in jet production.

In order to better constrain the magnetic flux, it is important to image the polarimetric structure at the BH shadow scales directly. For the M\,84 BH with an expected apparent size of $\approx 5$$\, \mu$as, this will be achieved with next-generation mm/sub-mm VLBI missions planned both on the ground and in space e.g., next-generation EHT \citep[ngEHT;][]{johnson2023key}. Such a study of M\,84 will provide a rare opportunity for bridging the gap in jet power between M\,87 (strong jet) and Sgr\,A$^*$ (absence of jet). 

\section{Summary} \label{sec:summary}
We presented the analysis of the jet collimation profile of M\,84 and an additional investigation of the core shift effect. The main conclusions of our studies are as follows:
\begin{itemize}
    \item In our study, we investigated the jet shape from $\sim 10^2\,r_{\rm s}$ up to $\sim 10^7\,r_{\rm s}$ and found the jet transition from a semi-parabolic to a conical shape. The inferred power-law indices are $a_{\rm u}=0.71 \pm 0.03$ and $a_{\rm d} = 1.16 \pm 0.01$ for the upstream and downstream jets, respectively.    
    \item The transition occurred at a location of $15.8 \pm 3.0$ mas, which is approximately $1.67 \times 10^4\,r_{\rm s}$ and almost two orders of magnitude smaller than the Bondi radius. This finding suggests that the transition at the Bondi radius is not universal and indicates a possible rapid decrease in the density profile of material around the central engine.
    \item We investigated the core shift effect in M\,84, which is crucial for determining a precise jet collimation profile. The core shift value falls in the range of $\lesssim 1$ mas between the lowest and highest frequencies (1.5 and 43\,GHz), following a best-fit value of $r_{\delta} \propto \nu^{-1/k}$, where $k= 0.78 \pm 0.23$. The asymptotic line leads to an upper limit estimation of the central black hole location at $\lesssim 14 \, r_{\rm s}$ relative to the 43\,GHz core.
    \item We derived the value of the magnetic field strength from the core shift effect. The resulting magnetic flux implies that MAD state could be realized in M\,84, which lies at the low end of LLAGNs.
    \item We compiled the transition distances and upstream power-law indices of M\,84 and previously studied LLAGNs. We found a hint of a correlation between the observed jet collimation parameter and normalized accretion rate, suggesting the importance of accretion activities in governing the LLAGN jet collimation mechanism.
\end{itemize}

\begin{acknowledgments}
We sincerely thank the anonymous referee for his/her careful reviewing that improved the manuscript. We thank R. Blandford, W. Baan, L. Petrov, K. Ebisawa, T. Kawashima, K. Toma, and M. Kino for valuable discussions. This research was supported by International Graduate Program for Excellence in Earth-Space Science (IGPEES), a World-leading Innovative Graduate Study (WINGS) Program, the University of Tokyo. K.H. and M.N. acknowledge financial support from JSPS KAKENHI (Grant 24K07100). K.H. also acknowledges support from MEXT/JSPS KAKENHI (Grants 21H04488 and 22H00157) and the Mitsubishi Foundation (Grants 201911019 and 202310034). Y.C. is supported by the Natural Science Foundation of China (Grant 12303021) and the China Postdoctoral Science Foundation (No. 2024T170845). M.H. acknowledges support from JSPS KAKENHI (Grant 19KK0081). W.J. thanks the support from the National Natural Science Foundation of China (Grant 12173074). VLBA and VLA are operated by the National Radio Astronomy Observatory, a facility of the National Science Foundation, under a cooperative agreement by Associated Universities, Inc.
\end{acknowledgments}

\bibliography{sample631}
\bibliographystyle{aasjournal}

\appendix

\section{Comparison of Standard and Atmosphere-Corrected Phase-Referencing Images of M\,84}\label{sec:app1}
\renewcommand{\thefigure}{Appendix I.\arabic{figure}}
\setcounter{figure}{0}

\begin{center}
\begin{figure}[h!]
  \centering
  \includegraphics[width=4.4cm]{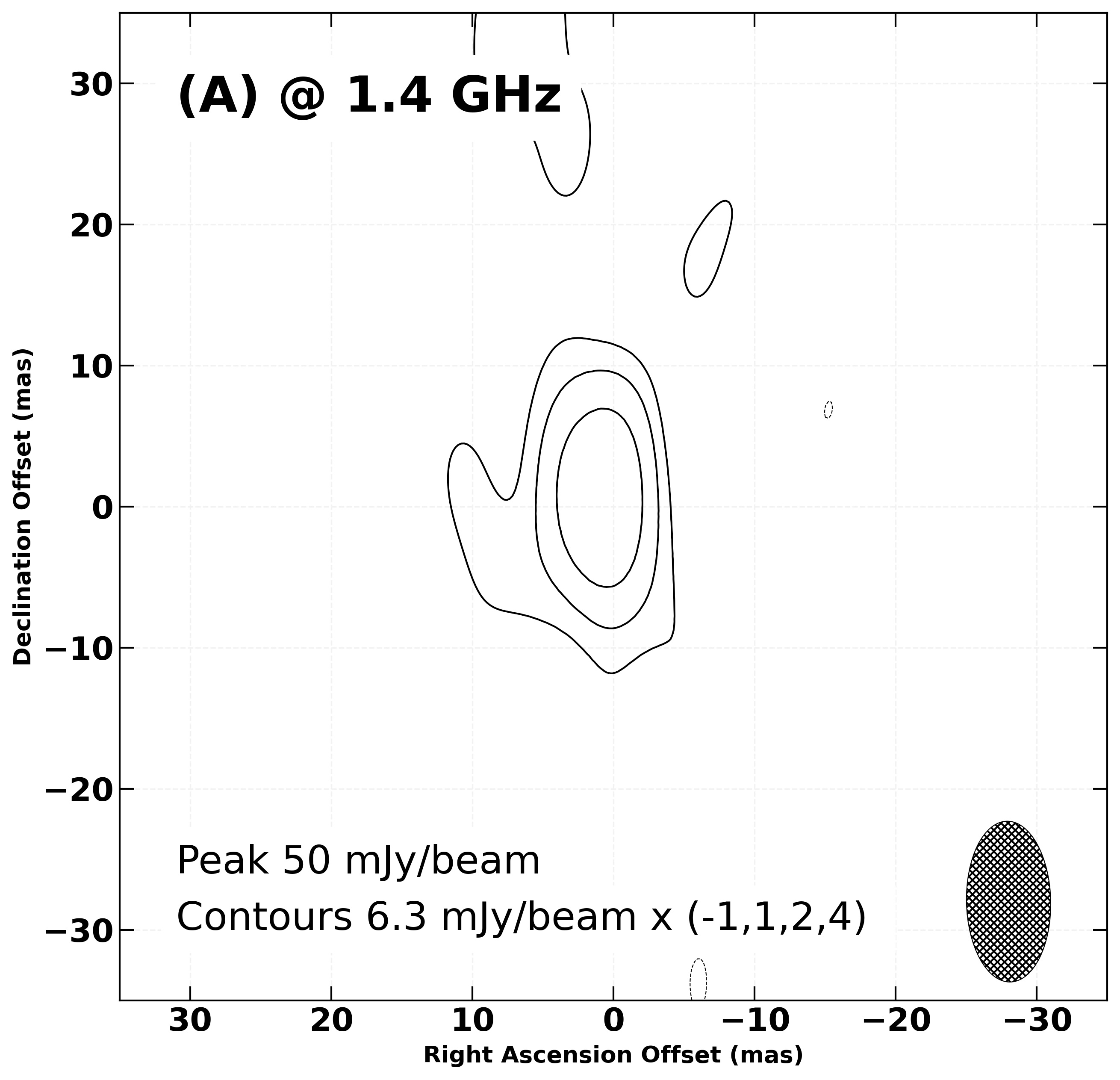}
  \includegraphics[width=4.4cm]{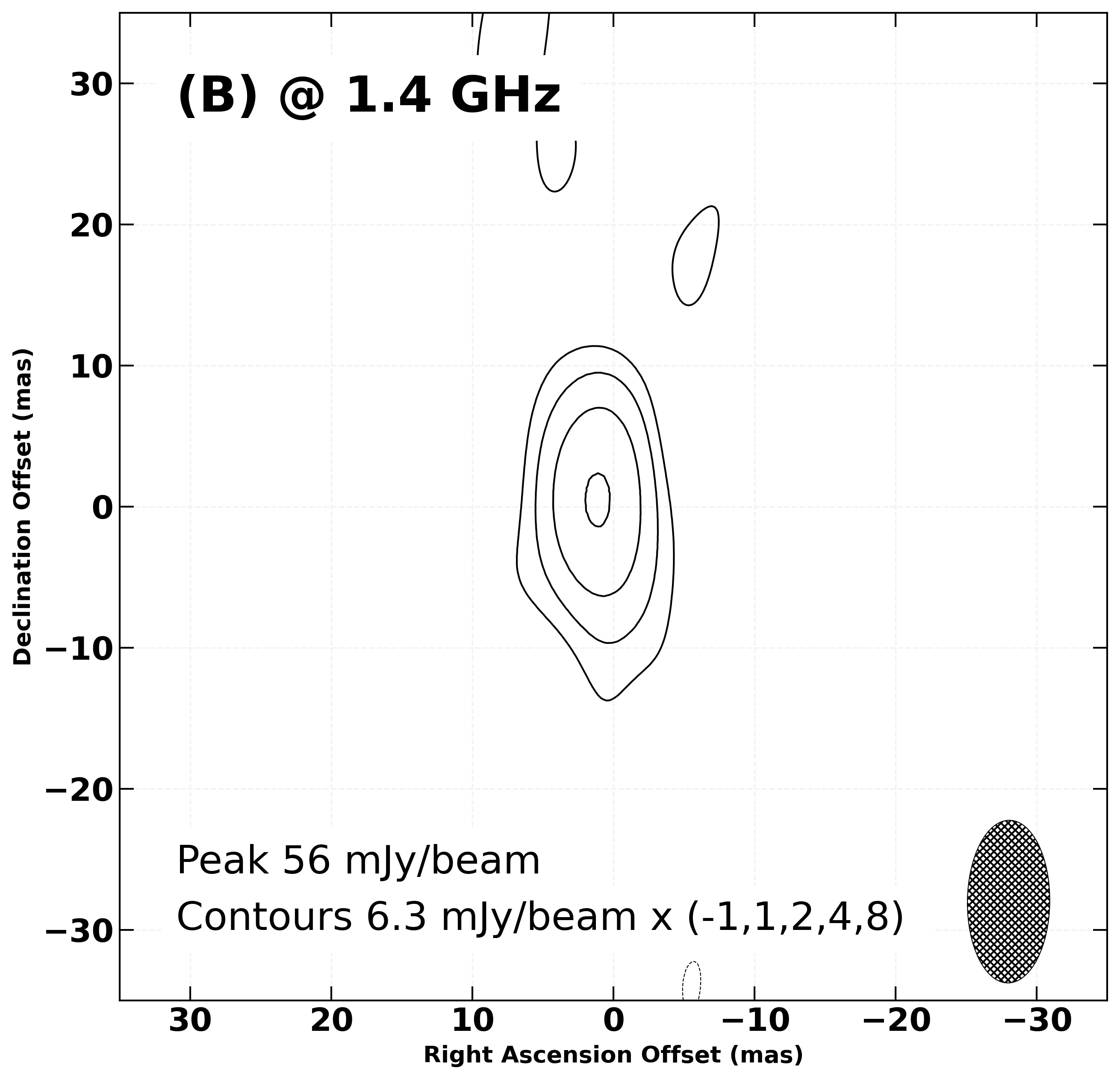}
  \includegraphics[width=4.4cm]{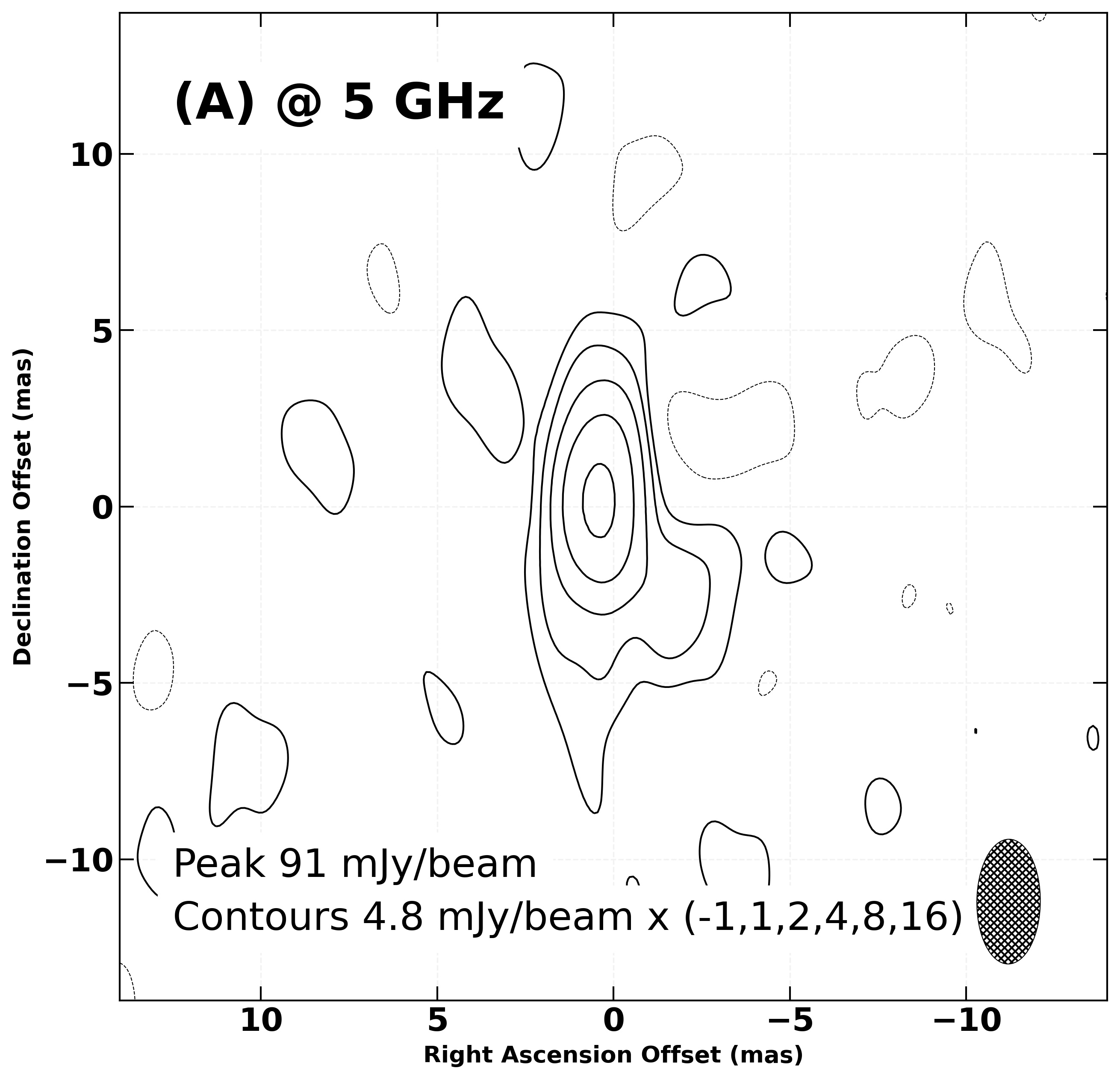}
  \includegraphics[width=4.4cm]{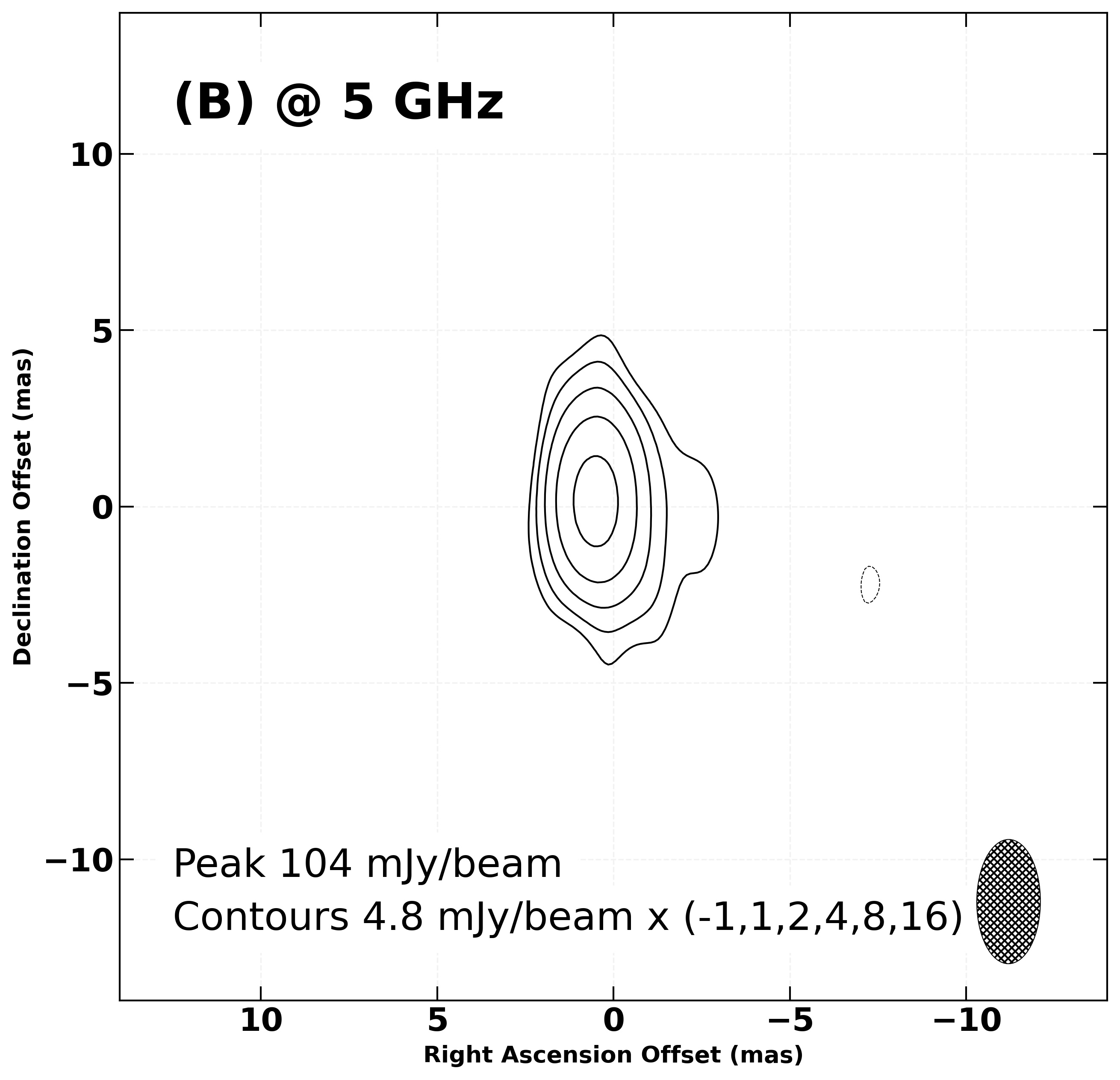}    \includegraphics[width=4.4cm]{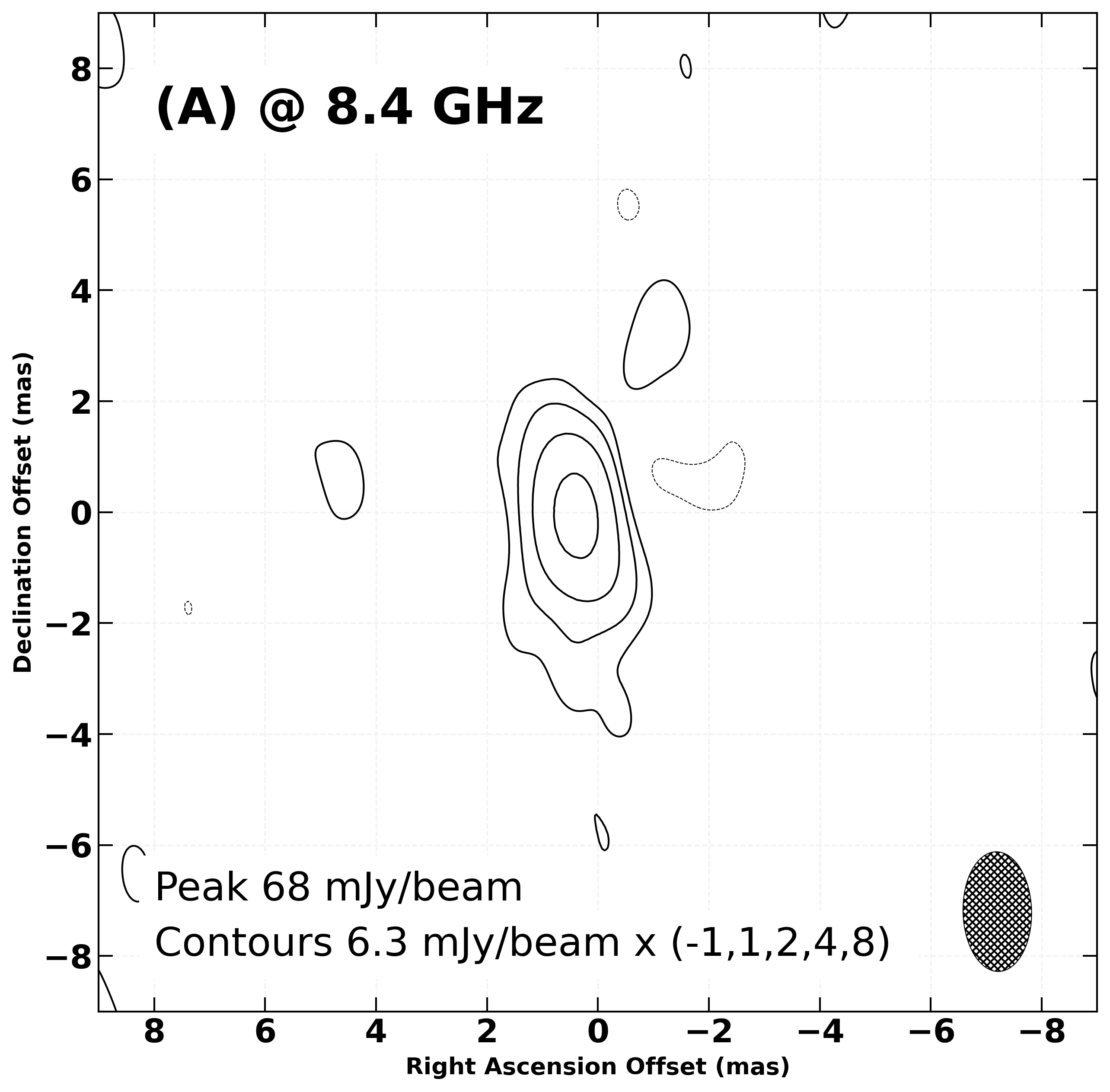}
  \includegraphics[width=4.4cm]{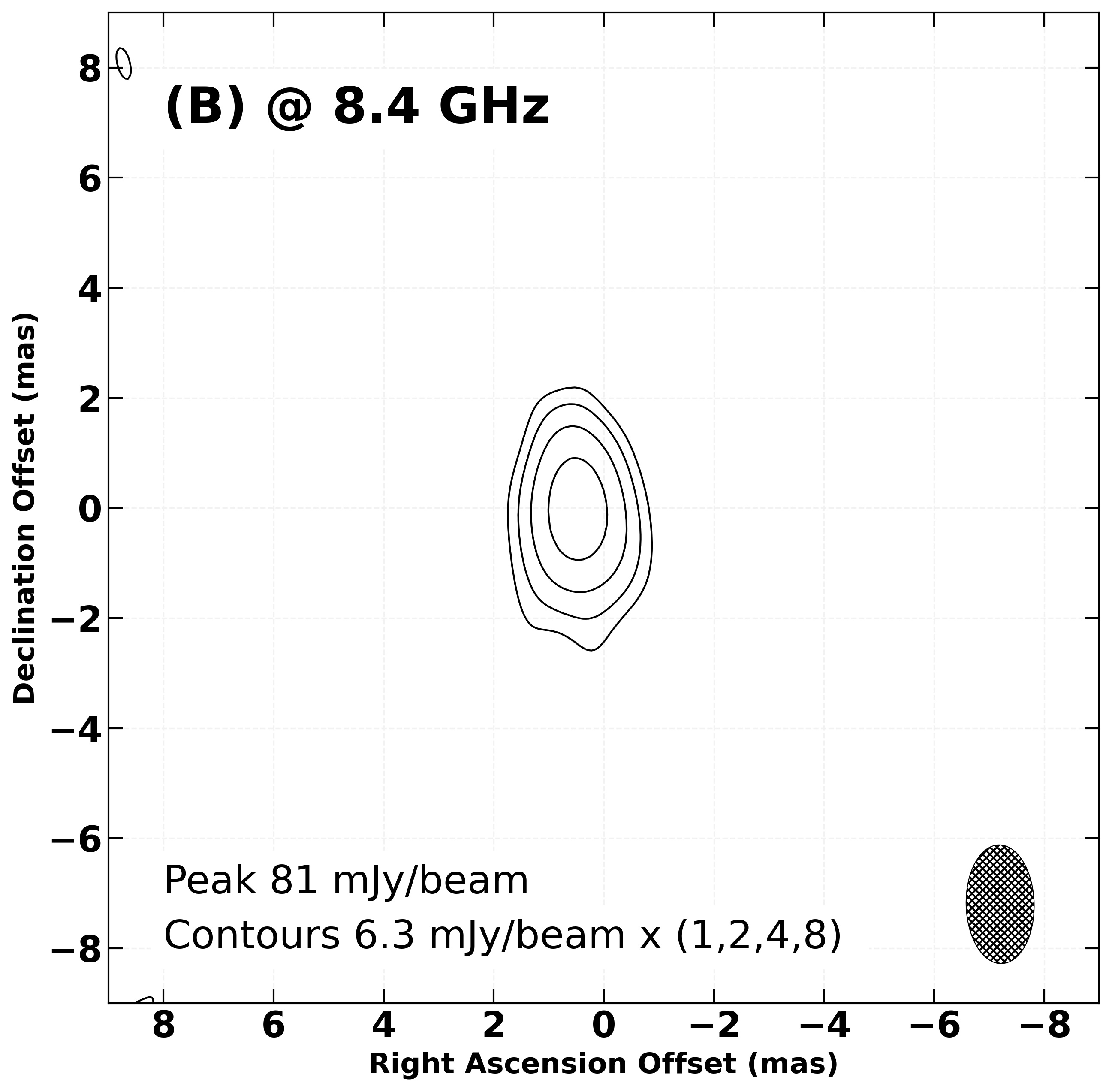}
  \includegraphics[width=4.4cm]{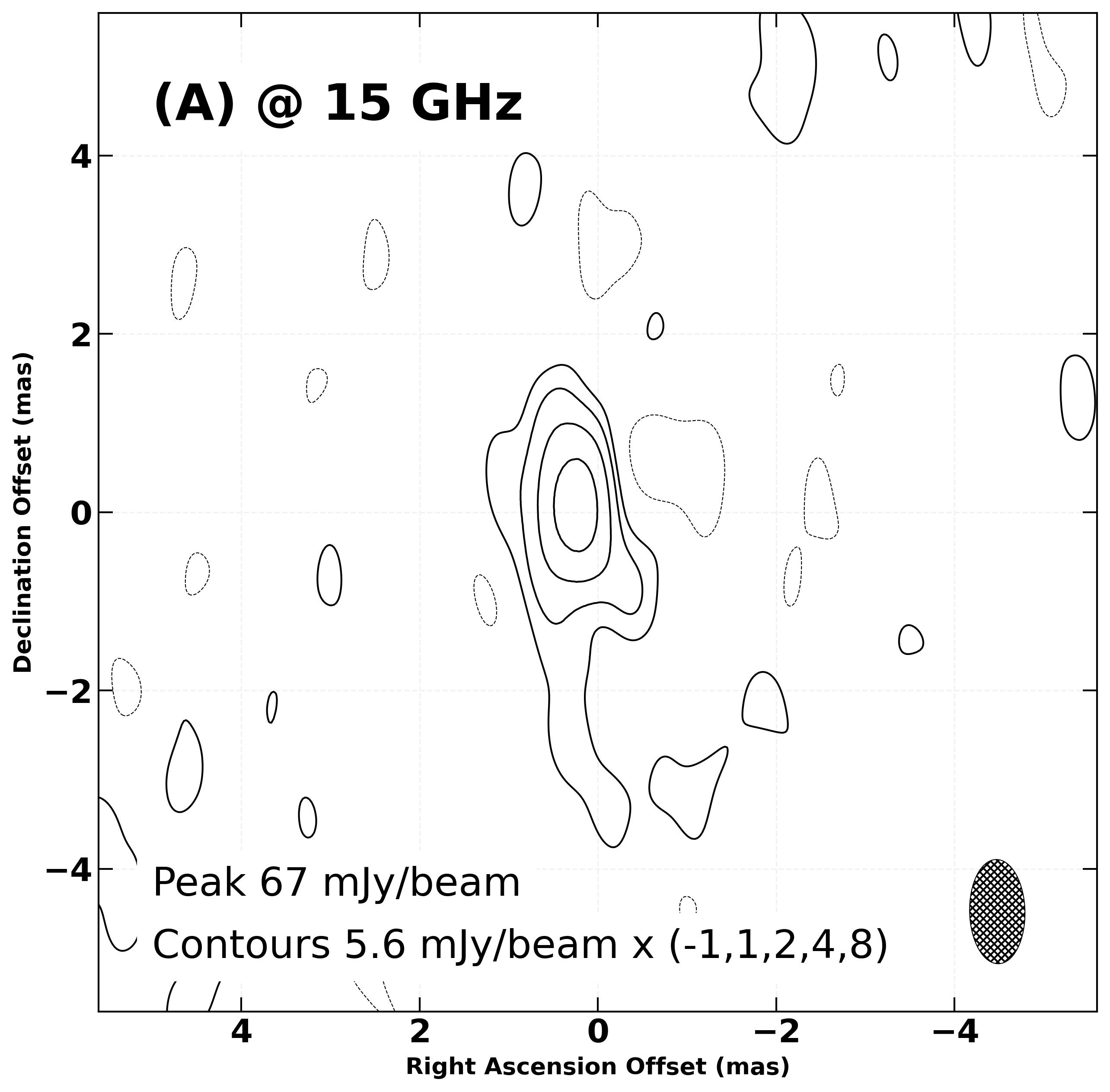}
  \includegraphics[width=4.4cm]{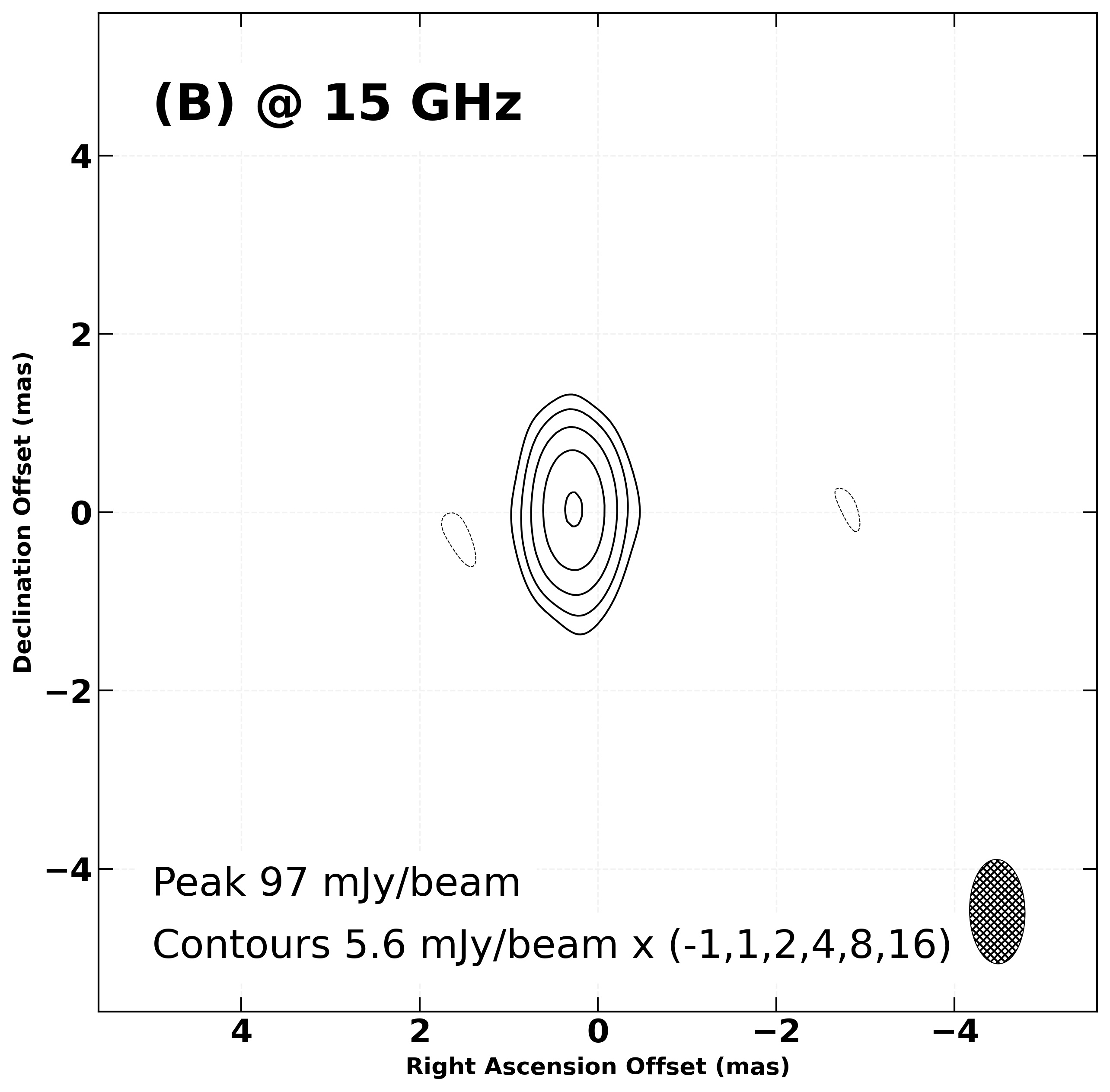}
    \includegraphics[width=4.4cm]{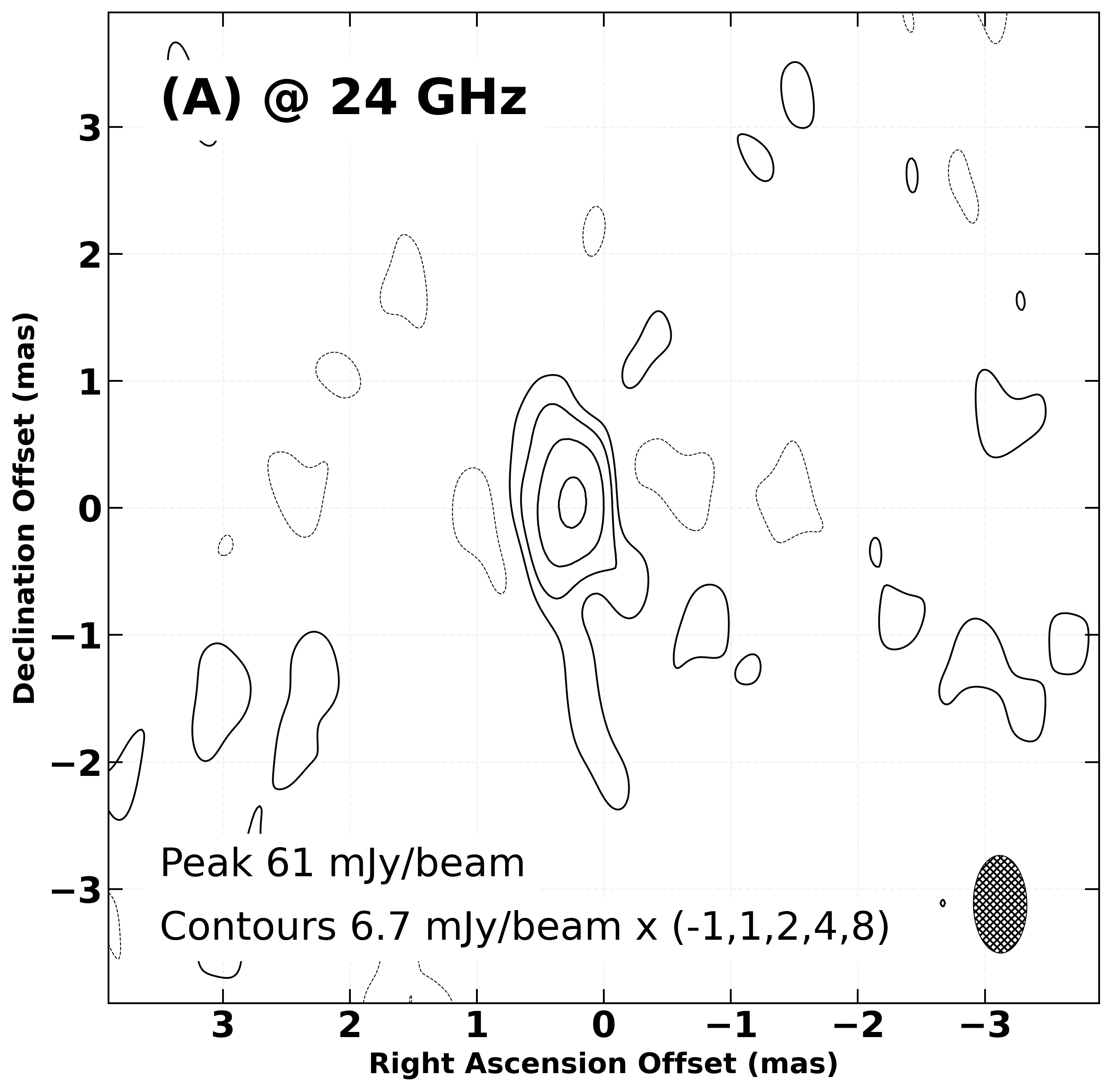}
  \includegraphics[width=4.4cm]{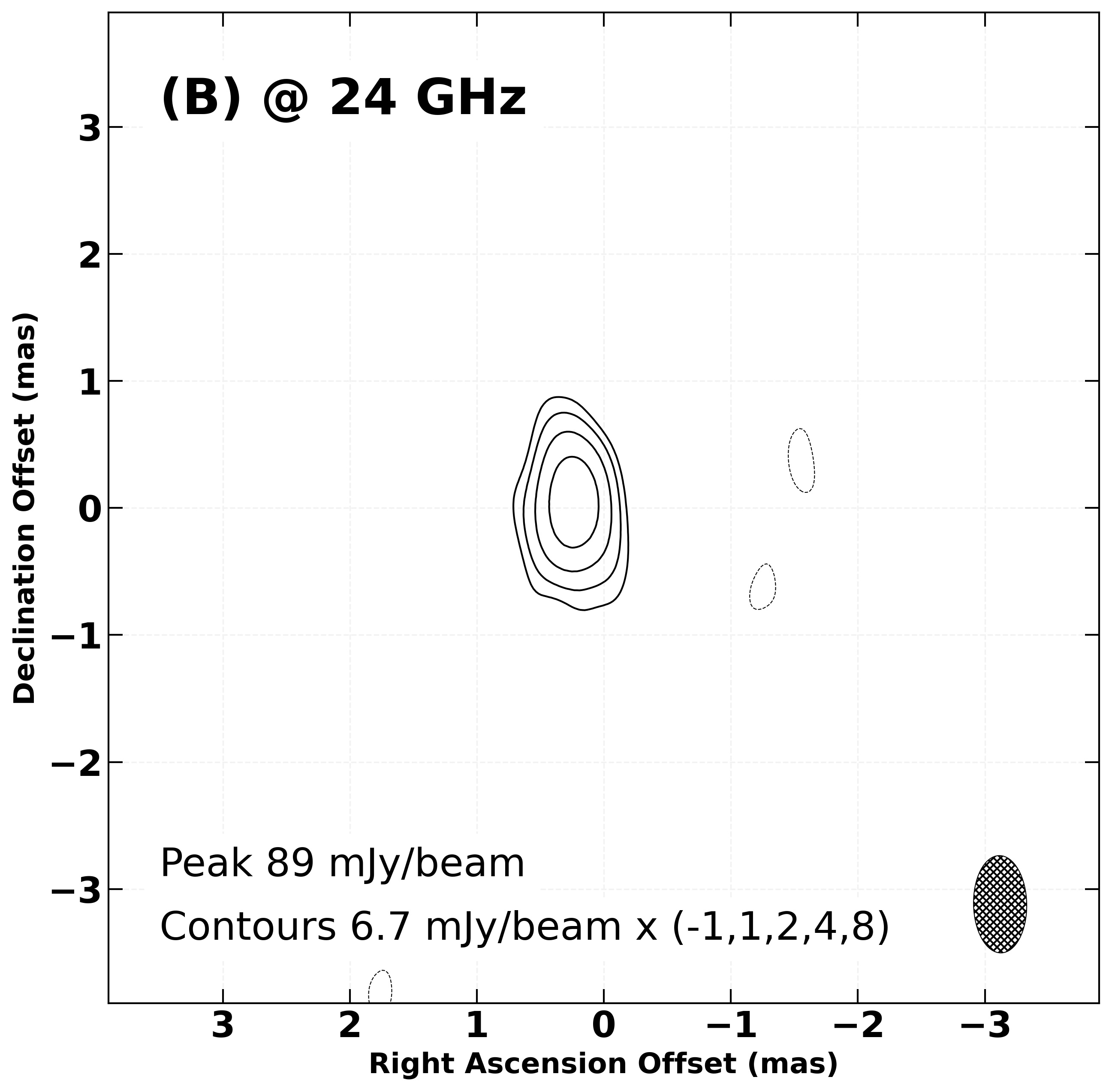}
  \includegraphics[width=4.4cm]{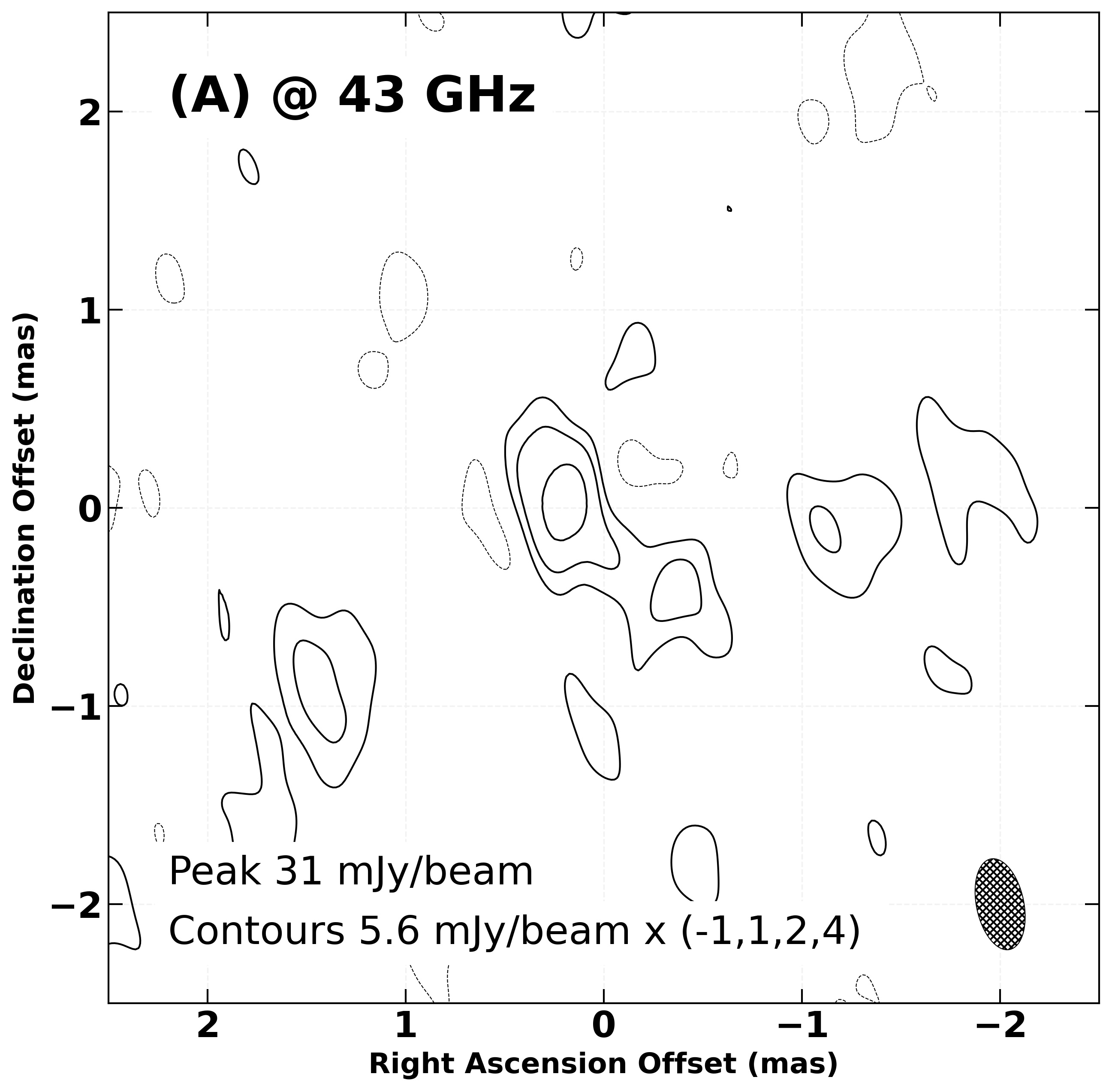}
  \includegraphics[width=4.4cm]{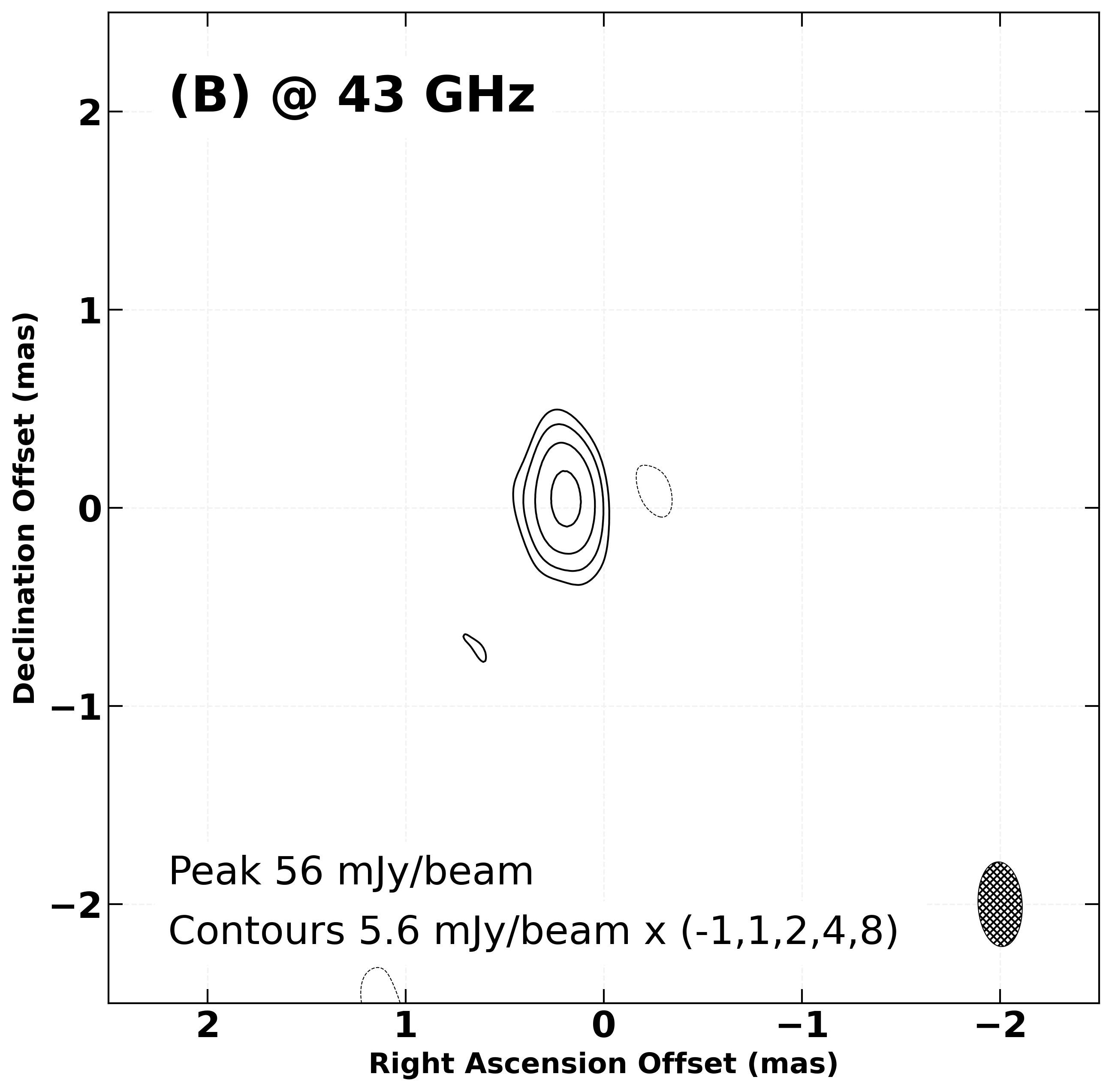}
  \caption{Standard phase-referencing (labeled as (A)) and atmosphere-corrected phase-referencing (labeled as (B)) images of M\,84 at 1.4, 5, 8.4, 15, 24 and 43\,GHz. At each frequency, the same contour levels are used in both images (A) and (B) for a better comparison, where the first contour starts from 3\,$\sigma$ of the image (B).}
\label{fig:PRSTR vs PRATM}
\end{figure}
\end{center}

In Figure\,\ref{fig:PRSTR vs PRATM}, we present a comparison of the standard and atmosphere-corrected phase-referencing images of M\,84 at 1.4, 5, 8.4, 15, 24 and 43\,GHz. At each frequency, the images (A) and (B) refer to the standard phase-referencing and atmosphere-corrected phase-referencing images, respectively. 
One can see the significant improvement of image dynamic range in the images (B) at all frequencies, underscoring the effectiveness of the corrections in mitigating atmospheric distortions, as discussed in Section \ref{sec:atm}. The improvement of peak-to-noise ratio is especially remarkable at 15\,GHz and higher (a factor of 2--3), where the non-dispersive tropospheric component is dominant. At the lower frequencies, the image improvement is relatively modest since the dispersive ionospheric residuals are more dominant than the tropospheric component that we corrected. Nevertheless, we confirm a factor of 1.3--1.9 image improvement also at 1.4, 5 and 8.4\,GHz.  
This enhancement allows for more precise measurements of the core position at all the observed frequencies, reinforcing the findings discussed in Section \ref{ssec:coreshift}. 

\newpage
\section{Parameters of LLAGNs}\label{sec:app2}
\renewcommand{\thetable}{Appendix II.\arabic{table}}
\setcounter{table}{0}

In this appendix, we list the black hole mass, Eddington accretion rate, Bondi accretion rate, and the electron density and temperature around the Bondi radius for the LLAGN objects referenced in Figure \ref{fig:LLAGNs}. These parameters are compared to assess their influence on the accretion rate and LLAGN characteristics, with the table summarizing key values for comparison.

\movetabledown=2.1in
\begin{rotatetable*}
\vspace*{1cm}
\begin{center}
    \small
    \begin{threeparttable}
    \caption{LLAGN Object Parameters\\}
    \renewcommand{\arraystretch}{1.2}
    \begin{tabularx}{22.8cm}{P{2cm}*{9}{Q{1.9cm}}}
    \hline
    \hline
     Source & $a_u$ & $r_0$ & $r_{\rm B}$& $M_{\rm BH}$ & $\dot{M}_{\rm E}$ & $\dot{M}_{\rm B}$ & $n_e (r_{\rm B})$\tnote{\textcolor{AASBlue}{3}} & $T (r_{\rm B})$\tnote{\textcolor{AASBlue}{3}} & References \\
    \rule{0pt}{3ex} & & $(r_{\rm s})$ & $(r_{\rm s})$ & $(M_{\odot})$ & $(M_{\odot}\, \rm yr^{-1})$ & $(M_{\odot}\,\rm yr^{-1})$ & $(\rm cm^{-3})$ & $(\rm keV)$ & 
    \\ \hline
     M\,84 & $0.72$ & $1.45 \times 10^4$ & $5.9 \times 10^5$ & $8.5 \times 10^8$ & 18.7 & $3.7 \times 10^{-3}$ & 0.27 & 0.73 & $[1]-[2]$
    \\ \hline
     M\,87 & $0.56-0.58$\tnote{\textcolor{AASBlue}{2}} & $2.5 \times 10^5$ & $4.0 \times 10^5$ & $6.5 \times 10^9$ & 143 & $3.5 \times 10^{-1}$ & 0.23 & 0.52 & $[3]-[7]$
    \\ \hline
     NGC\,315 & $0.45-0.58$ & $(3-42) \times 10^3$ & $7.5 \times 10^5$ & $2.08 \times 10^9$ & 45.8 & $5 \times 10^{-2}$ & 0.28 & 0.44 & $[8]-[12]$
    \\ \hline
     NGC\,4261 & $0.56-0.62$ & $(4-5) \times 10^3$ & $6.5 \times 10^5$ & $1.62 \times 10^9$ & 35.6 & $4.5 \times 10^{-2}$ & 0.17 & $0.63$  & $[13]-[17]$
    \\ \hline
     NGC\,6251 & $0.5$\tnote{\textcolor{AASBlue}{2}} & $2.1 \times 10^5$ & $4 \times 10^5$ & $6 \times 10^8$& 13.2 & $5 \times 10^{-2}$ & $0.1$ & $0.6$ & $[18]-[21]$
    \\ \hline
     NGC\,1052 & $0.18-0.5$ & $(3-11) \times 10^3$ & $5.4 \times 10^5$ & $1.5 \times 10^8$ & 3.4 & $3.4 \times 10^{-2}$ & 0.18 & 0.62 & $[22]-[27]$
    \\ \hline
    \end{tabularx}
    \tablecomments{References: [1] \cite{bambic2023agn}, [2] \cite{walsh2010supermassive}, [3] \cite{asada2012structure}, [4] \cite{hada2013innermost},  [5] \cite{allen2006relation}, [6] \cite{akiyama2019first}, [7] \cite{russell2013radiative}, [8] \cite{park2021jet}, [9] \cite{boccardi2021jet}, [10] \cite{boizelle2021black}, [11] \cite{ricci2022exploring}, [12] \cite{worrall2007inner}, [13] \cite{nakahara2018finding}, [14] \cite{yan2023kinematics},  [15] \cite{sawada2022circumnuclear}, [16] \cite{gliozzi2003origin}, [17] \cite{worrall2010jet}, [18] \cite{tseng2016structural}, [19] \cite{ferrarese1996evidence}, [20] \cite{migliori2011implications}, [21] \cite{gliozzi2004xmm}, [22] \cite{baczko2022ambilateral}, [23] \cite{baczko2024putative}, [24] \cite{woo2002active}, [25] \cite{lo2023surveying}, [26] \cite{artyukh2009magnetic}, [27] \cite{osorio2020inner}.}
    \footnotesize
      \item[2] The power-law index value from \citet{asada2012structure} and \citet{tseng2016structural} was originally expressed in the form of $W \propto r^{(1/a)}$ and has been converted to $W \propto r^{a}$ for consistency.
      \item[3] Note that M\,84 and M\,87 are two of only five known systems in which the Bondi radius is resolved by \textit{Chandra}, despite its sub-arcsecond angular resolution \citep{bambic2023agn}. For the other LLAGN objects, electron density and temperature are estimated within the nuclear region on the order of 1 arcsecond.
    \end{threeparttable}
    \label{tab:LLAGNtable}
\end{center}
\end{rotatetable*}

\end{document}